\documentclass[twocolumn,useAMS,usenatbib]{mn2e}
\usepackage{graphicx,amsmath}
\usepackage{bm}

\newcommand{\rmd}{{\rm d}}
\newcommand{\rme}{{\rm e}}
\newcommand{\rmi}{{\rm i}}

\newcommand{\ds}{\ensuremath{\Delta\Sigma}}
\newcommand{\hmpc}{$h^{-1}$Mpc}
\newcommand{\hkpc}{$h^{-1}$kpc}

\newcommand{\scinv}{\ensuremath{\Sigma_c^{-1}}}
\newcommand{\MD}{\mathbfss D}
\newcommand{\MJ}{\mathbfss J}
\newcommand{\MR}{\mathbfss R}
\newcommand{\MT}{\mathbfss T}
\newcommand{\arctanh}{\,{\rm arctanh}\,}
\newcommand{\eint}{e^{(0)}}
\newcommand{\eints}[1]{e^{(0)\,#1}}

\newcommand{\combo}[2]{\left(\!\!\begin{array}{c}#1\\#2\end{array}\!\!\right)}

\newcommand{\beq}{\begin{equation}}
\newcommand{\eeq}{\end{equation}}
\newcommand{\beqa}{\begin{eqnarray}}
\newcommand{\eeqa}{\end{eqnarray}}

\title[Density profiles]
{Density profiles of galaxy groups and clusters from SDSS galaxy-galaxy 
weak lensing}

\author[Mandelbaum et al.]
 {Rachel Mandelbaum$^1$\thanks{Electronic address:
    {\tt rmandelb@princeton.edu}},
  Uro\v s Seljak$^{1,2}$, 
  Richard J. Cool$^3$, Michael Blanton$^4$, 
\newauthor
Christopher M. Hirata$^5$, Jonathan Brinkmann$^6$
\\$^1$Department of Physics, Jadwin Hall, Princeton University,
      Princeton, NJ 08544, USA
\\$^2$International Centre for Theoretical Physics, Strada Costiera 11,
      34014 Trieste, Italy
\\$^3$Steward Observatory, 933 N. Cherry Avenue, Tucson AZ 85721, USA
\\$^4$New York University, Center for Cosmology and Particle Physics,
 4 Washington Place, New York NY 10003, USA
\\$^5$Institute for Advanced Study, Einstein Drive, Princeton, NJ 08540, USA
\\$^6$Apache Point Observatory, 2001 Apache Point Road,
      Sunspot NM 88349, USA
}

\date{\today}

\begin{document}
\maketitle

\begin{abstract}
We present results of a measurement of the shape of the density
profile of galaxy groups and clusters 
traced by 43~335 Luminous Red Galaxies (LRGs) with
spectroscopic redshifts from the Sloan Digital Sky Survey (SDSS).  The
galaxies are selected such that they are the brightest within a
cylindrical aperture, split into two luminosity samples,  and modeled
as the sum of stellar and dark matter 
components. We present a detailed investigation of many possible systematic 
effects that could contaminate our signal and develop methods to remove them, 
including a detected intrinsic alignment 
for galaxies within 100\hkpc\ of LRGs which we remove using photometric 
redshift information. The resulting lensing signal is consistent with NFW 
profile dark matter halos; the SIS profile is ruled out at the 96
(conservatively) and 99.96 per cent confidence level (CL) for the
fainter and brighter  
lens samples (respectively) when we fit using lensing data between 40 
\hkpc\ and $2$ \hmpc\ with total signal-to-noise of $19$ and $25$ for the 
two lens samples. The lensing signal amplitude suggests that the faint 
and bright sample galaxies typically reside in haloes of mass $(2.9\pm 
0.4)\times 10^{13}h^{-1}M_{\odot}$ and $(6.7\pm 0.8)\times 10^{13} 
h^{-1}M_{\odot}$ respectively, in good agreement with predictions based on 
halo spatial density with normalization lower than the 'concordance'
$\sigma_8=0.9$.  When fitting for the concentration parameter in the NFW 
profile, we find $c = 5.0 \pm 0.6 ({\rm stat}) \pm 1 ({\rm sys})$, and $c = 
5.6 \pm 0.6 ({\rm stat}) \pm 1 ({\rm sys})$ for the faint and bright
samples, consistent with $\Lambda$CDM simulations.  We also split the
bright sample further to determine masses and concentrations for
cluster-mass halos, finding mass $(1.3\pm 0.2)\times 10^{14}
h^{-1}M_{\odot}$ for the sample of LRGs brighter than $-22.6$ in r. 
We establish that on average there is a correlation between 
the halo mass and central galaxy 
luminosity relation that scales as $M \propto L^2$. 
\end{abstract}

\begin{keywords}
gravitational lensing --- galaxies: haloes.
\end{keywords}

\section{Introduction}

Elliptical galaxies, particularly Luminous Red Galaxies (LRGs),
are good tracers of the most massive galactic halos.  They have a
number of useful properties, including simple, well-understood spectral energy
distributions that can be modeled as a passively-evolving burst of
star formation, with a constant low level of ongoing star
formation; variation with environment that is quantified as well
\citep{2001AJ....122.2267E,2003MNRAS.344.1000B,2003ApJ...585..694E};
uniform photometric properties, including the 
fundamental plane relating luminosity, radius, and central velocity
dispersion
\citep{1973ApJ...179..731F,1977ApJ...216..214V,1987ApJ...313...59D,1989ARA&A..27..235K,1992MNRAS.254..601B,1994ARA&A..32..115R,2003AJ....125.1866B}
of which the Faber-Jackson relation \citep{1976ApJ...204..668F} is a projection; and
simple light profiles, the de Vaucouleurs 
profile, with $I\propto \exp(-r^{1/4})$.  Galaxies on the
massive end of the red sequence contain the majority of the stellar
mass of the universe
\citep{1998ApJ...503..518F,2002AJ....124..646H,2004ApJ...608..752B,2004ApJ...600..681B,2006AJ....131..736C},
and hence their properties are of great
interest. It was shown that while elliptical galaxies dominate over
spirals in overdense environments, these galaxies do reside in a wide
variety of environments, from the field to the richest clusters 
(\citealt{1975ApJ...199..545M}; 
\citealt{1977ApJ...211..309A};
\citealt{1994Natur.369..462P}; \citealt{1999ApJ...520L...1V}; Loh
\& Strauss, 2006, {\it in prep.}), with the 
majority of the LRGs located in group to small cluster sized halos.  
There have also
been extensive studies of LRG clustering
\citep{2005ApJ...619..178E,2005ApJ...633..560E,2005astro.ph.12166M,2005ApJ...621...22Z}
from 10~\hkpc\ to 100~\hmpc\ scales. 

In addition to the extensive photometric studies of elliptical
galaxies already in the literature, we would like to add a good
understanding of their density profiles.  This information is
useful for several purposes. First, the dark matter profile can give
us information that can be related to the formation history and
merger history of these galaxies.  Second, the shape of dark matter (DM)
profiles can be predicted using N-body simulations
\citep{1996ApJ...462..563N,1997ApJ...477L...9F,1997ApJS..111...73K,1998ApJ...499L...5M,1999MNRAS.310..527A,2000ApJ...544..616G,2000ApJ...529L..69J,2001MNRAS.321..559B,2001ApJ...554..903K,2001ApJ...557..533F,2002ApJ...568...52W,2003ApJ...588..674F,2003ApJ...597L...9Z,tasitsiomi_etal04,2005MNRAS.364..665D},
though with some
disagreement in the value of inner asymptotic logarithmic slope,
possibly due to resolution issues, and hence LRGs in massive halos 
provide a particularly good way of testing the density profile predictions of
$\Lambda$CDM.  The brightest red galaxies have been shown to have
small baryon fractions compared to fainter red galaxies and compared
to typical blue galaxies (\citealt{2005astro.ph.11164M}), so the effects of
baryons on the DM profile are relatively small in this class of
galaxies.  Furthermore, since the virial radii are so large, and the
concentration parameters decrease with mass, any changes in
logarithmic slope around the scale radius occur at large scales
relative to smaller galaxies, and hence are easier to measure using
the tools available.

While the matter distributions in elliptical galaxies can be simply
probed on small scales (within a few $r_e$) using central velocity dispersions
\citep{2000A&AS..144...53K,2002MNRAS.329..513D,2004NewA....9..329P},
finding probes of the dark matter profiles 
on larger scales is more challenging.  Kinematic tracers such as
satellite galaxies can give information out to tens of \hkpc.
Hydrostatic analyses of X-ray intensity profiles, and 
strong- and weak-lensing constraints for individual clusters, are
numerous but thus far do not give a single clear picture
\citep{1998MNRAS.296..392A,2000ApJ...539..540C,2001A&A...379..384C,2001ApJ...549L..33R,2001ApJ...554..881S,2002A&A...384..743A,2002ApJ...572...66A,2003ApJ...583..606K,2004ApJ...600L...7H,2005ApJ...623L...5F,2005astro.ph..7092V,2006ApJ...636..698F,2006astro.ph..1301H,2006astro.ph..1628K}.
Weak lensing is 
another tool that can be used out to scales of several \hmpc, and the
SDSS provides a particularly advantageous dataset due to its
significant sky coverage and spectroscopic redshifts (which allow us
to determine the profile as a function of transverse separation).
\cite{2001ApJ...548L...5H} and \cite{2005ApJ...634..806P} have done a
group lensing analysis with a much smaller sample ($\sim 100$) groups from
CNOC2; however, these small samples are not sufficient to allow a
detailed determination of the shape of the density profile.

Measurements of the 3-dimensional density profiles can be used to learn about
cosmology as well.  For example, the concentration parameter of the
profile can be related to the halo mass via the dark matter power
spectrum normalization $\sigma_8$ and the matter density $\Omega_m$.
Thus, measuring the profile for several samples of varying halo mass
can teach us about the underlying cosmology.

We begin in \S\ref{S:theory} with a description of the theory behind
the measurement we are making. \S\ref{S:data} has a full description
of the dataset used, including the properties of the lens and source
samples, and possible causes of systematic error.  We present results
of profile measurements in \S\ref{S:results}, and conclude in
\S\ref{S:conclusion}. 

Here we note the cosmological model and units used in this paper.
All computations assume a flat $\Lambda$CDM universe with
$\Omega_m=0.3$, $\Omega_{\Lambda}=0.7$, and $\sigma_8=0.9$.  Distances
quoted for 
transverse lens-source separation are comoving (rather than physical)
\hkpc, where $H_0=100\,h$ km$\mathrm{s}^{-1}\,\mathrm{Mpc}^{-1}$.
Likewise, \ds{} is computed using the expression for \scinv{} in
comoving coordinates, Eq.~\ref{E:sigmacrit}.  In the units
used, $H_0$ scales out of everything, so our results are independent of
this quantity.  Finally, 2-dimensional separations are indicated with the
capital $R$, 3-dimensional radii with lower-case $r$ (occasionally $r$
may denote $r$-band magnitude as well).

\section{Theory}\label{S:theory}

Here we describe the theory behind our attempts to measure the
three-dimensional density profile using the galaxy-galaxy weak lensing
signal. 

\subsection{Lensing}

Galaxy-galaxy weak lensing provides a simple way to probe the
connection between galaxies and matter via their
cross-correlation function
\beq
\xi_{g,m}(\vec{r}) = \langle \delta_g (\vec{x})
\delta_{m}^{*}(\vec{x}+\vec{r})\rangle 
\eeq
where $\delta_g$ and $\delta_{m}$ are overdensities of galaxies and
matter, respectively.  This cross-correlation can be related to the
projected surface density
\beq\label{E:sigmar}
\Sigma(R) = \overline{\rho} \int \left[1+\xi_{g,m}\left(\sqrt{R^2 + \chi^2}\right)\right] d\chi
\eeq
(where $r^2=R^2+\chi^2$) which is then related to the observable
quantity for lensing, 
\beq\label{E:ds}
\ds(R) = \gamma_t(R) \Sigma_c= \overline{\Sigma}(<R) - \Sigma(R), 
\eeq
where the second relation is true only in the weak lensing limit, for
a matter distribution that 
is axisymmetric along the line of sight (which is naturally achieved
by our procedure of stacking thousands of lenses to determine their
average lensing signal).  This observable quantity can
be expressed as the product of two factors, a tangential shear
$\gamma_t$ and a geometric factor
\beq\label{E:sigmacrit}
\Sigma_c = \frac{c^2}{4\pi G} \frac{D_S}{D_L D_{LS}(1+z_L)^2}
\eeq
where $D_L$ and $D_S$ are angular diameter distances to the lens and
source, $D_{LS}$ is the angular diameter distance between the lens
and source, and the factor of $(1+z_L)^{-2}$ arises due to our use of
comoving coordinates.  For a given lens redshift,
$\Sigma_c^{-1}$ rises from zero at $z_s = z_L$ to an asymptotic value
at $z_s \gg z_L$; that asymptotic value is an increasing function of
lens redshift.  

For this paper, we are primarily interested in the contribution to the
galaxy-mass cross-correlation from the galaxy halo profile itself
(central galaxy Poisson term),
rather than from neighboring halos (halo-halo term), and hence 
\beq\label{E:sigmar2}
\Sigma(R) = \int_{-\infty}^{\infty} \rho(r=\sqrt{\chi^2+R^2}) d\chi
\eeq
The halo-halo term for galaxies in host halos can be modeled simply
using the galaxy-dark matter cross-power spectrum as in, e.g.,
\cite{2005MNRAS.362.1451M}, and is only important for $R > 2$ \hmpc,
which is the maximum scale probed in this paper.  

\subsection{Profiles}\label{SS:profiles}

In previous analyses \citep{2005MNRAS.362.1451M,2005astro.ph.11164M},
we have modeled the lensing 
signal as the sum of contributions from central galaxies (those that
lie in host halos) and from satellite galaxies (those that lie in
subhalos).  For this work, to simplify interpretation by
avoiding issues such as tidal stripping of satellites and the radial
distribution of satellites within host halos, we use methods
(described in \S\ref{SS:lenses}) that enable us to isolate a sample of central
galaxies with negligible contamination from satellites.  Thus, for
this work, our interpretation is much
simpler, and can be related directly to $\rho(r)$ for the lens sample
under consideration.

While more complicated models may be useful for techniques
capable of resolving smaller scale information than this work (minimum
transverse separation of 20 \hkpc), we model each galaxy as the sum of
two components.  The first component is a dark matter profile, with
the following density profile:
\begin{equation}\label{E:genprofile}
\rho(r) = \frac{\rho_s}{\left(r/r_s\right)^{-\alpha}
  \left(1+r/r_s\right)^{\alpha+\beta}}. 
\end{equation}
which gives a logarithmic slope of
\begin{equation}\label{E:logslope}
\frac{\rmd\ln{\rho}}{\rmd\ln{r}} = \alpha - (\alpha+\beta) \frac{r/r_s}{1+r/r_s},
\end{equation}
or $\alpha$ for $r\ll r_s$ and $-\beta$ for $r\gg
  r_s$.  This profile with $\alpha=-1$ and $\beta=3$ is the NFW profile
  \citep{1996ApJ...462..563N}, and this shape (with varying values of
  $\alpha$) is a generic prediction of
  $\Lambda$CDM N-body simulations over a large range of masses.  In
  addition to the logarithmic slopes, it can be  defined by two  
  parameters, $c=r_{vir}/r_s$ and $M$.  The virial radius $r_{vir}$
  and $\rho_s$ can be 
  related to $M$ via consistency relations.  The first is that the
  virial radius is that within which the average density is equal to
  $180\overline{\rho}$:
\begin{equation}\label{E:rvir}
M = \frac{4\pi}{3}r_{vir}^3 \left(180\overline{\rho}\right)
\end{equation}
Note that this definition differs from the oft-used $200\rho_{crit}$
definition of mass by roughly 30 per cent for typical values of
concentration.    The second relation, 
used to determine $\rho_s$ from $M$ and $c$, is simply that the
volume integral of the density profile out to the virial radius must
equal the virial mass (though when computing the signal, we do not
truncate the profiles beyond $r_{vir}$).  In principle, the NFW 
$c$ is a weakly decreasing function of halo mass, with a variety of expressions
used \citep{2001MNRAS.321..559B,2001ApJ...554..114E}, making this
profile a one-parameter family of 
profiles, but in order to give generality to the fits,
we allow it to be a free parameter.

To generate the profile at a particular redshift, we start with a
given mass, use Eq.~\ref{E:rvir} to get the comoving virial radius
using the comoving matter density, so that $M$,  $\alpha$, $\beta$,
and $c$ define the profile uniquely.  When computing the lensing
signal, we are 
actually averaging over a range of redshifts, which means that since
we define the profile in comoving coordinates, we 
expect that the measured profile will be somewhat blurred due to evolution of
concentration with redshift.

While the more general ellipsoidal versions of these density profiles
are seen universally in $\Lambda$CDM simulations, here we work only with
the spherical versions, because our process of averaging over thousands of
lens galaxies makes the results insensitive to halo
triaxiality.  This statement is not made on the basis of
  analytical proofs (and surely must be wrong if one requires an
  extremely high level of precision), but rather on the basis of the
  lensing signal measured
  in N-body simulations as an average over randomly-oriented
triaxial NFW profiles.  In \cite{2005MNRAS.362.1451M}, it was shown
that spherical NFW profiles do an excellent job at describing the
lensing signal for a variety of luminosity bins, with the masses and
concentrations of the best-fit profiles related to the real masses and
concentrations in the simulations in a particular way (to be discussed
further below).  The fit $\chi^2$ values were very good when the
errorbars used on the signal were one-tenth of our current errorbars,
so henceforth we consider only spherical NFW fits rather than trying
to account for the averaging of triaxial halos.

The second component to the full $\rho(r)$ that we consider is the
baryon density.  This includes not only the stellar mass, but also any
baryons that did not form stars.  \cite{2005astro.ph..2517R}
estimate that 80 per cent of the baryonic mass in galaxies
is in the form of stars, though for elliptical galaxies this number is
closer to 100 per cent, so there is very little gas that has not been
converted to stars.  As is commonly done for
ellipticals, we model the $\rho(r)$ for the stellar component as a
Hernquist profile
since it is a deprojected de Vaucouleurs profile.  The Hernquist
profile can be expressed via Eq.~\ref{E:genprofile} with $\alpha=-1$
and $\beta=4$ (and hence falls off faster than an NFW profile on large
scales).  In the absence of further information about any ``dark''
baryonic component, we do not model its profile separately, 
with the expectation that the need for a dark baryonic component will
be shown by poor fits in the inner regions.

Finally, we cannot in reality assume that these density profiles are
independent.  The gravitational potential due to one will necessarily
affect the form of the other, so the dark matter profile will not be
expected to conform exactly to the ones seen in N-body simulations on
small scales where baryonic matter is significant.  This effect is
known as adiabatic contraction (AC, \citealt{1986ApJ...301...27B},
\citealt{2004ApJ...616...16G}, 
\citealt{2005astro.ph.10539G}, \citealt{2005ApJ...634...70S}) and has been
studied with analytic models and simulations.  To model adiabatic
contraction, we use the {\sc contra}\footnote{\slshape
http://www-astronomy.mps.ohio-state.edu/\~{}ognedin/contra/} code 
from
\cite{2005astro.ph.10539G} 
to obtain adiabatically-contracted density profiles $\rho(r)$.
Generally, on the scales we consider, which are several times the de
Vaucouleurs radius, we will show that AC leads to an immeasurably
small change in the profile, and hence is not necessary for this work.
This is because of low baryonic fractions in these galaxies.

\begin{figure}
\includegraphics[angle=0,width=3.2in]{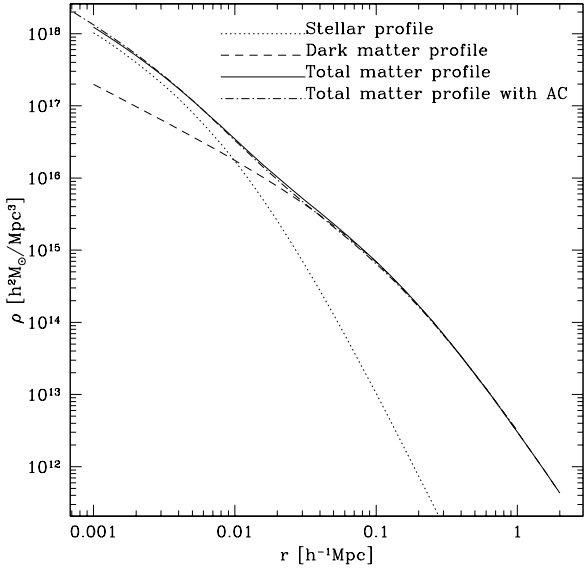}
\caption{\label{F:rho}Density profile $\rho(r)$ for the typical
  LRG described in the text, including the profile separately for
  dark matter and for baryons, the total profile, and the total
  profile including AC (barely distinguishable from the profile
  without AC), as labelled on the plot.} 
\end{figure}

Figure~\ref{F:rho} shows the density profile $\rho(r)$ for a typical
expected LRG
profile, with $M_{NFW}=6\times 10^{13} h^{-1}M_{\odot}$, $M_b = 8\times 10^{11}
h^{-1}M_{\odot}$, $c=7$, baryonic $r_s=10$ \hkpc, $\alpha=-1$, $\beta=3$,
$\Sigma_c^{-1}=2h^{-1}\times 10^{-4} pc^2 M_{\odot}^{-1}$, with and
without AC.  The source for these numbers are the predicted LRG masses
from the halo mass function, a calculation that will be described
shortly.

\begin{figure}
\includegraphics[angle=0,width=3.2in]{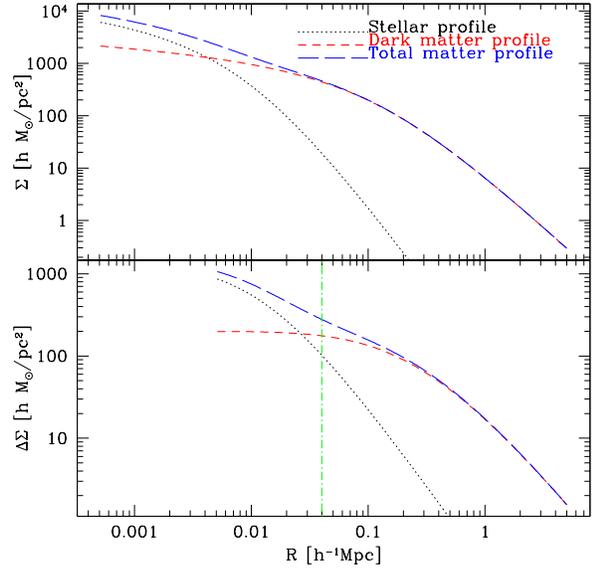}
\caption{\label{F:plotsigma}Surface density $\Sigma(R)$ and projected
  surface density $\Delta\Sigma(R)$ for the typical LRG described in
  the text, including the profile separately for
  dark matter and for baryons, and the total profile without AC (AC
  produces a negligible difference at the relevant transverse
  separation).  A vertical line at $40$ \hkpc shows the minimum scale
  used for the fits.} 
\end{figure}

Given that we measure the lensing signal for projected separations
$R>20$ \hkpc, and fit using $R>40$ \hkpc, one might wonder why a
baryonic component is necessary for our modeling at all, given the
values of the density profiles in Fig.~\ref{F:rho} at these
separations. We present Fig.~\ref{F:plotsigma} to answer this
question.  The top panel of this figure shows the projected surface
density $\Sigma$ as a function of transverse separation for the same
density profile as in 
Fig.~\ref{F:rho}, and the bottom panel shows the surface density
contrast, $\Delta\Sigma$, which is the quantity of interest for
lensing.  As shown in the top panel of Fig.~\ref{F:plotsigma}, the
baryonic component produces a significant increase in the surface density for
transverse separations $R<\sim 20$ \hkpc. However, what is important
for the computation of $\Delta\Sigma$ is that also leads to a
significant {\it steepening} of $\Sigma$ on these scales.  This
steepening of $\Sigma$ leads to a noticable increase in $\Delta\Sigma$
even on larger scales due to the dependence on 
$\Delta\Sigma(<R)$, out to $100$ \hkpc{} for this 'typical'
profile.  This effect is the reason why our modeling requires a
baryonic component in $\Delta\Sigma$ for $20<R<100$ \hkpc\ despite its
relatively small contribution to $\rho$ for similar scales.

As mentioned above, the full lensing signal for host galaxies
(Eq.~\ref{E:sigmar}) is 
the sum of the terms due to the cross-correlation between a host
galaxy and its own halo (Eq.~\ref{E:sigmar2}), and the halo-halo term
(cross-correlation with the dark matter in other halos). 
We must still consider
the halo-halo term.  Simple modeling of the h-h term as in
\cite{2000MNRAS.318..203S} indicates that below $2$ \hmpc, 
this term is $\ll \Delta\Sigma$ due to the central profile.

One factor that we must consider when relating the best-fit profile
parameters to reality is the effect of averaging over a distribution
of halo masses.  We start with the halo mass function, $dn/dM$ (which
depends on redshift), to specify the distribution of halo masses:
\beq
\frac{dn}{dM}dM = \frac{\overline{\rho}}{M}f(\nu)d\nu
\eeq
where $\overline{\rho}$ is the mean matter density of the universe.
The function $f(\nu)$ can we written in a universal form when defined
in terms of peak height
\beq
\nu = \left[ \frac{\delta_c}{\sigma(M)}\right]^2
\eeq
where $\delta_c=1.686$ is the linear overdensity at which a spherical
perturbation collapses, and $\sigma(M)$ is the rms fluctuation in
spheres containing on average mass $M$ at an initial time,
extrapolated using linear theory to $z$. We use the 
\cite{1999MNRAS.308..119S} version of the mass function, 
\beq
\nu f(\nu) = A(1+\nu'^{-p})\sqrt{\nu'}e^{-\nu'/2}
\eeq
where $\nu'=a\nu$, with $a=0.707$ and $p=0.3$.  The original
\cite{1974ApJ...187..425P} mass function corresponds to $a=1$ and
$p=0$.  The constant $A$ is determined by mass conservation, requiring
\beq
\int_{0}^{\infty} f(\nu)d\nu = 1.
\eeq

Once we have normalized the mass function, we can compare the comoving
density of the LRG sample to the predicted density of halos with mass
above some minimum mass $M_{min}$ to estimate minimum halo masses,
assuming that the majority of these massive halos are populated by red
rather than blue galaxies.  For a comoving density of $\overline{n}_{LRG}= 1.2\times
10^{-4} (h/\mbox{Mpc})^3$, as for the full LRG sample, the
characteristic sample redshift suggests that
\beq
\int_{M_{min}}^{\infty} \frac{dn}{dM} dM = \overline{n}_{LRG}
\eeq
for $M_{min} \sim 10^{13.4} = 2.5 \times 10^{13} h^{-1}M_{\odot}$.  As
will be shown, we also split the sample into two, such that the
fainter sample contains roughly 2/3 of the galaxies.  We can model
this split by saying the fainter luminosity sample has the same
$M_{min}$ and includes all halos up to some $M_1$ (i.e., a top-hat
filter on $n(M)$), using the fact that light roughly traces mass for
these galaxies.  Then, we find $M_{1}$ using
\beq
\int_{M_{min}}^{M_1} \frac{dn}{dM} dM = \frac{2}{3}\overline{n}_{LRG}
\eeq
which gives $M_{1} \sim 10^{13.8} = 6 \times 10^{13}
h^{-1}M_{\odot}$.  This $M_1$ is then the minimum halo mass for the
brighter LRG sample containing $1/3$ of the LRGs.

%
For these two samples containing a range of halo masses, we must then
consider the relationship between $\Delta\Sigma$ averaged over $n(M)$,
i.e.
\beq
\langle \Delta\Sigma\rangle = \frac{\int \Delta\Sigma
  \frac{dn}{dM}dM}{\frac{dn}{dM} dM}
\eeq
over the relevant mass range ($M_{min}$ to $M_1$, or $M_1$ to
$\infty$).
Because of differences in expected concentration parameters, one might
expect that this averaging process will change not only the signal
amplitude, but also its shape, possibly changing best-fit
concentrations and slopes.  

 The properties of
these distributions, and the results of performing our fits to the
averaged signals, are found in Table~\ref{T:dndmresults}.  The
important lesson is that the averaging process has little effect on
the best-fit concentration parameter, as one would hope.  The average
signal can be modeled as an NFW profile to a high degree of accuracy.
\begin{table*}
\caption{\label{T:dndmresults}Approximate values of minimum, maximum, mean,
  and median predicted masses for the two LRG samples using the halo
  mass function $dn/dM$ for two cosmologies, and the best-fit masses
  to the model signal.} 
\begin{tabular}{lccccccc}
\hline\hline
Sample & $M_{min}$ & $M_{max}$ & $\langle M\rangle$ & $M_{median}$ &
$\langle c\rangle$ & $M_{nfw,fit}$ & $c_{nfw,fit}$ \\
 & $10^{13}h^{-1}M_{\odot}$ & $10^{13}h^{-1}M_{\odot}$ &
$10^{13}h^{-1}M_{\odot}$ & $10^{13}h^{-1}M_{\odot}$ & &
$10^{13}h^{-1}M_{\odot}$ & \\ 
\hline
\hline
\multicolumn{8}{c}{$\Omega_m=0.3$, $\sigma_8=0.9$}\\
Faint LRGs & $2.5$ & $6.3$ & $3.8$ & $3.5$ & $7.4$ & $3.8$ & $7.3$ \\
Bright LRGs & $6.3$ & $\infty$ & $14$ & $9.8$ & $6.3$ & $13$ & $6.1$ \\
\multicolumn{8}{c}{$\Omega_m=0.25$, $\sigma_8=0.75$}\\
Faint LRGs & $1.6$ & $3.2$ & $2.2$ & $2.0$ & $6.5$ & $2.2$ & $6.2$ \\
Bright LRGs & $3.2$ & $\infty$ & $7.1$ & $4.9$ & $5.5$ & $6.0$ & $5.0$
\\
\hline\hline
\end{tabular}
\end{table*}
We remind the reader, however, that the process we have described
incorporates only the averaging over a distribution in halo masses,
but not any scatter in the mass-luminosity relationship or the
concentration-mass relationship.  As shown in
\cite{2005MNRAS.362.1451M}, the scatter in these
relationships can have more profound effects on the signal than the
simple average over distributions.

Another averaging process that may concern us is the average over
redshifts.  Because the nonlinear mass changes with redshift, the
predicted value of concentration parameter $c$ also changes with
redshift.  Hence, this may cause some blurring of the profile that
will affect best-fit parameters.  
However, since our average over halo masses is also an average over
concentration parameters, and was found to have little effect on the
profile shape, and because the change of concentration with $z$ is
small, this effect is also not expected to be significant.

\subsection{Method}

Our method is to fit physically-motivated models
with the sum of a stellar plus dark matter component to \ds.  In
practice, these 
6-parameter fits, with four parameters for the dark matter and two for
the baryonic profile, are 
highly degenerate due to the scarcity of information about small
scales, so in order to obtain results, we make physically-motivated
assumptions.  For each
sample, we do the following fits in order of increasing complexity:
\begin{enumerate}
\item \label{L1}Fix to a one-parameter fit, a singular isothermal
  sphere (SIS), and fit only for the mass.
\item \label{L2}Fix to a more general single power-law profile,
  fitting for the mass and the logarithmic slope.
\item Fix $\alpha$, $\beta$, and the stellar component; fit for $c$
  and NFW mass.
\item Again, fix to an NFW profile, but try fitting for
  other profile parameters such as inner or outer logarithmic slope.
\end{enumerate}

One fact that may help us in these fits is that
$\Delta\Sigma$ to some degree mixes information about $\rho$ at
different scales.  For given transverse separations $R$,
$\Delta\Sigma(R)$ includes information from lower values of $r$ due
to the averaging $\overline{\Sigma}(<R)$.

\section{Data}\label{S:data}

The data used here are obtained from the SDSS
\citep{2000AJ....120.1579Y}, an ongoing survey to image roughly
$\pi$ steradians of the sky, and follow up approximately one million of
the detected objects spectroscopically \citep{2001AJ....122.2267E,
2002AJ....123.2945R,2002AJ....124.1810S}. The imaging is carried out
by drift-scanning the sky 
in photometric conditions \citep{2001AJ....122.2129H,
2004AN....325..583I}, in five bands ($ugriz$) \citep{1996AJ....111.1748F,
2002AJ....123.2121S} using a specially-designed wide-field camera
\citep{1998AJ....116.3040G}. These imaging data are the source of the
Large-Scale Structure (LSS)
sample that we use in this paper. In addition, objects are targeted for
spectroscopy using these data \citep{2003AJ....125.2276B} and are observed
with a 640-fiber spectrograph on the same telescope
\citep{2006AJ....131.2332G}. All of these data are 
processed by completely automated pipelines that detect and measure
photometric properties of objects, and astrometrically calibrate the data
\citep{2001adass..10..269L, 2003AJ....125.1559P,mtpipeline}. The SDSS is well
underway, and has had six major data releases \citep{2002AJ....123..485S,
2003AJ....126.2081A, 2004AJ....128..502A, 2005AJ....129.1755A,
2004AJ....128.2577F, 2006ApJS..162...38A}. 

\subsection{Lenses}\label{SS:lenses}

The galaxies used as lenses are those targeted as the spectroscopic
Luminous Red Galaxy (LRG) sample
\citep{2001AJ....122.2267E}, including area beyond Data Release 4 (DR4).
The total area coverage is 5154 square degrees, as
available in the most recent version of the NYU Value Added Galaxy
Catalog (VAGC, \citealt{2005AJ....129.2562B}).

We include these galaxies in the
redshift range 
$0.15<z<0.35$, where the upper cutoff is designed to ensure that the
lenses still have sufficient number of sources behind them.  Within
these redshift limits, the sample is approximately volume-limited with
a number density of $1.2\times 10^{-4} (h/\mbox{Mpc})^3$,
once we apply the additional color-magnitude cuts described in Loh \&
Strauss (2006, {\it in prep.})\footnote{We thank Yeong-Shang Loh for
  providing the necessary files to implement this cut.} to eliminate
contamination from 
fainter, bluer galaxies below $z\sim 0.23$. (We note that this  is the
number
density before we make our cut requiring that the galaxy be a host galaxy;
the numbers in Table~\ref{T:lenses} give a lower average number density.)

\begin{figure}
\includegraphics[angle=0,width=3.4in]{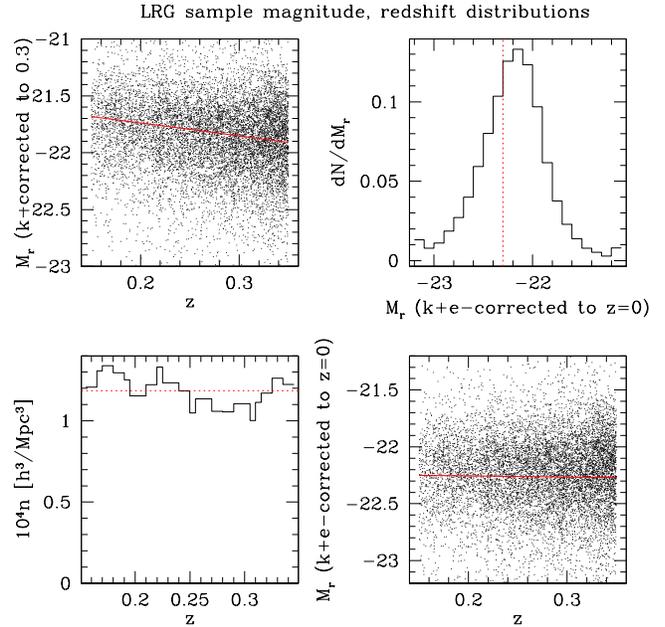}
\caption{\label{F:magz}Upper left panel: scatter plot of absolute magnitude in $r$
  band, k-corrected to $z=0.3$, versus redshift; line shows average trend with
  redshift (total change of 0.2 magnitudes across the full redshift range).  Upper right: absolute magnitude distribution $dn/dM_r$
  when k+e-corrected to $z=0$, with a line indicating the cut value
  between the faint and bright bins.  Lower right: $M_r$ k+e-corrected to
  $z=0$ versus redshift; line shows average trend with redshift, which
  is nearly constant (total change of 0.02 magnitudes across the
  full redshift range).  Lower left: comoving number density as a
  function of redshift, with a line indicating the average value.} 
\end{figure}

Model (not Petrosian) magnitudes were used for all luminosity cuts
described in this paper; in all cases, $h=1$ magnitudes were used, so
the division into samples does not depend on $h$.  Magnitudes are corrected
for extinction using 
reddening maps from \cite{1998ApJ...500..525S}.  We
apply a k+e-correction (combined k-correction and correction for
evolution of the spectrum) to all magnitudes to redshift zero as in
Wake, et. al. (2006, {\it in prep.})
using the \cite{2003MNRAS.344.1000B}
stellar population synthesis code\footnote{We thank David Wake for
  providing the k+e-corrections for the appropriate models.}.  
We note that if we do not use these k+e-corrections, but simply
use {\sc kcorrect v4\_1\_4} \citep{2003AJ....125.2348B} to the sample
median redshift of 0.27, the resulting absolute
magnitude histogram suggests that the galaxies above the median
redshift are on average brighter than those below it.  This result is
consistent with the sign of passive evolution, for which galaxies are
brighter in the past.  In order to avoid our luminosity cuts selecting
intrinsically different samples at high and low redshift, we apply all
luminosity cuts using the k+e-corrected luminosities.
Figure~\ref{F:magz} shows information about the redshift and magnitude
distribution of the sample. 

In order
to study variation with luminosity, we divide the sample into two
subsamples at $M_r = -22.3$; roughly 2/3 of the sample is fainter than
this value, and 1/3 is brighter, giving approximately equal signal to
noise in each bin.  Sample parameters are included in
Table~\ref{T:lenses}. 

Scale radii for the baryonic components were obtained by finding the
average de Vaucouleurs profile fit radius for each sample from {\sc
  Photo}, the SDSS processing pipeline; all radii quoted here are
half-light radii.

For this paper, we isolate ``host'' versus ``satellite galaxies''
using spectroscopic galaxy counts in cylinders of comoving radius 2 
$h^{-1}$Mpc and line-of-sight length $\Delta v = \pm 1200$
km~s$^{-1}$.  Since we only want to eliminate
galaxies for which there is another, brighter spectroscopic LRG
nearby, we simply require that those LRGs in our ``host'' sample
either (a) be the only one in the cylinder, or (b) be the brightest in
the cylinder.

A problem with this scheme is fiber collisions; since fibers cannot be
placed closer than 55'', corresponding to 200 \hkpc{} at these
redshifts, we may in principle still have some satellite galaxies
that will distort the observed profile shape.  Note that roughly 9 per
cent of spectroscopic LRGs lack redshifts due to either fiber
collisions, or due to redshift failure.  We address this issue by
using {\it all} galaxies passing the spectroscopic LRG sample cut I
(below $z=0.36$) and the additional stringent color-magnitude cut from
Loh \& Strauss (2006, {\it in prep.}), whether or not a spectrum was
obtained, in order to 
test whether a galaxy is a host or a satellite.  For those
LRGs without spectra, we used the redshift of the nearest LRG if there
was one within 55''; otherwise, we used photometric redshifts from
{\sc Kphotoz v4\_1\_4} with luminous red galaxy templates
\citep{2003AJ....125.2348B}, with 
$\sigma_z \sim 0.03$ for these bright red galaxies.  If the galaxy
without a spectrum turned out to be brighter than a galaxy with a
spectrum within a cylinder, we use the brighter one despite the lack
of a spectrum.   This cut on local environment reduces the sample
  by roughly 8 per cent.

\begin{table*}
\caption{\label{T:lenses}The lens samples used for this analysis,
  including redshift, absolute magnitude, stellar mass limits, and
  typical scale radii.}
\begin{tabular}{lcccccccc}
\hline\hline
Sample & $N$ & $z$ limits & $\langle z\rangle$ & $z_{eff}$ & $M_r$ limits & $\langle
L/L_{\odot}\rangle$ & $\langle M_*\rangle$ & $r_{deV}$ \\
 & & & & & & $h^{-2}10^{10}$ & ($10^{11}h^{-1}M_{\odot}$) & $h^{-1}$kpc \\
\hline
Faint LRGs & 27~700 & $[0.15, 0.35]$ & 0.27 & 0.24 &
$M_r > -22.3$ & 5.2  & 2.2 & 8 \\
Bright LRGs & 15~635 & $[0.15, 0.35]$ & 0.27 & 0.24 &
$M_r \le -22.3$ & 8.6  & 3.7 & 11 \\
\hline\hline
\end{tabular}
\end{table*}

We obtained stellar mass estimates in the following way.  First, we
use the estimates from 
\cite{2004NewA....9..329P}, which found that $M_{stellar}/L \sim
3M_{\odot}/L_{\odot}$ for ellipticals (this is also consistent with
the relationship between $M_{stellar}$ and $L$ seen in
\citealt{2005astro.ph.11164M} for bright red galaxies). 
The average luminosities of the faint and bright LRG samples 
determined in terms of $L_{\odot}$ using the absolute magnitude of
stellar luminosity determined in \cite{2003ApJ...592..819B} are
$5.2$ and $8.6h^{-2}\times 10^{10} L_{\odot}$ respectively, and with
$L_*=1.2h^{-2}\times 
10^{10}L_{\odot}$, we find (with appropriate factors of $h=0.7$)
stellar masses of $2.2$ and $3.7\times 10^{11} h^{-1}M_{\odot}$,
respectively.  

%
Finally, we will further split the ``bright'' LRG sample at
$M_r=-22.6$ in order to better trace the $M(L)$ relation.  When this
split is done, the $M_r<-22.6$ sample contains $1/9$ of the full LRG
sample, and the $-22.3 > M_r > -22.6$ sample contains $2/9$.

\subsection{Sources}

The source sample used is the same as that originally described in
\citet{2005MNRAS.361.1287M}, hereinafter M05.  This source sample
includes over 30 million galaxies from the SDSS imaging data with
$r$-band model magnitude brighter than 21.8, with
shape measurements obtained using the REGLENS pipeline, including PSF
correction done via re-Gaussianization \citep{2003MNRAS.343..459H} and
with cuts designed to avoid various shear calibration biases.  A full
description of this pipeline can be found in M05; here we include only
a brief summary.

The REGLENS pipeline obtains galaxy images in the $r$ 
and $i$ filters from the SDSS ``atlas images'' 
\citep{2002AJ....123..485S}.  The basic principle of shear measurement 
using these images is to fit a Gaussian profile with elliptical isophotes 
to the image, and define the components of the ellipticity
\beq
(e_+,e_\times) = \frac{1-(b/a)^2}{1+(b/a)^2}(\cos 2\phi, \sin 2\phi),
\label{eq:e}
\eeq
where $b/a$ is the axis ratio and $\phi$ is the position angle of the 
major axis.  The ellipticity is then an estimator for the shear,
\beq
(\gamma_+,\gamma_\times) = \frac{1}{2\cal R}
\langle(e_+,e_\times)\rangle,
\eeq
where ${\cal R}\approx 0.87$ is called the ``shear responsivity'' and 
represents the response of the ellipticity (Eq.~\ref{eq:e}) to a small 
shear \citep{1995ApJ...449..460K, 2002AJ....123..583B}.  In practice, a 
number of corrections need to be applied to obtain the ellipticity.  The 
most important of these is the correction for the smearing and 
circularization of the galactic images by the PSF; 
\citet{2005MNRAS.361.1287M} uses the PSF maps obtained from stellar images 
by the {\sc psp} pipeline \citep{2001adass..10..269L}, and corrects
for these 
using the re-Gaussianization technique of \citet{2003MNRAS.343..459H},
which includes corrections for non-Gaussianity of both the galaxy
profile and the PSF.  In order that these corrections can be successful,
we require that the galaxy be well-resolved compared to the PSF in
both $r$ and $i$ bands (the only ones used for shape measurement),
where we define the Gaussian resolution factor
\beq\label{E:R2def}
R_2 = 1-\frac{T^{(P)}}{T^{(I)}}
\eeq
where the $T$ values are the traces of the adaptive covariance
matrices, and the superscripts indicate whether they are of the PSF or
of the galaxy image.

M05 includes a
lengthy discussion of shear calibration biases in this catalog;
we will only summarize these issues briefly here.  Our source sample
is divided into three subsamples: $r<21$, $r>21$, and high-redshift
LRGs \citep{2001AJ....122.2267E}, defined
using color and magnitude cuts as in M05 using selection 
criteria related to those from \cite{2001AJ....122.2267E} and
\cite{2005MNRAS.359..237P}.  Using simulations 
from \citet{2003MNRAS.343..459H} to estimate the PSF dilution
correction and analytical models for selection biases and other issues
that affect shear calibration, we place $2\sigma$ limits on the shear
calibration bias of $[-0.05, 0.12]$ for $r<21$, $[-0.08, 0.18]$ for
$r>21$, and $[-0.06, 0.19]$ for LRGs.

As shown in Eq.~\ref{E:ds}, the lensing signal $\ds$ is a product of
the shear and factors involving lens and source redshifts.  Since the
lenses have spectroscopic redshifts, the primary difficulty is
determining the source redshift distribution.  We take three
approaches, all described in detail in M05.
For the $r<21$ sources, we use photometric redshifts and their error
distributions determined using a sample of galaxies in the Groth strip
with redshifts from DEEP2 \citep{2003SPIE.4834..161D,2003ApJ...599..997M,2004ApJ...609..525C,2005ASPC..339..128D}, and require
$z_s>z_l+0.1$ to avoid contamination from physically-associated
lens-source pairs.  For the $r>21$ sources, we use redshift
distributions from DEEP2.  For the high-redshift LRGs, we use
photometric redshifts and their error distributions determined using
data from the 2dF-Sloan LRG and Quasar Survey (2SLAQ), and presented in
\cite{2005MNRAS.359..237P}.  Note that the LRG lenses used in this paper 
are typically at higher redshift than the lenses used in M05 and for this 
reason we are more sensitive to errors in the source redshift 
distribution.  We will 
discuss the implications of this in \S\ref{ss:amplitude}.

Finally, we have placed constraints on other issues affecting the 
calibration of the lensing signal, such as the sky subtraction problem, 
intrinsic alignments, magnification bias, star-galaxy separation, and 
seeing-dependent systematics.  As shown in M05 the calibration of the 
signal using the three source samples agrees to within 10 per cent, with a 
total $1\sigma$ calibration uncertainty estimated at 7 per cent ($r<21$) 
or 10 per cent ($r>21$ and LRG).

\subsection{Signal computation}

Here we describe the computation of the lensing signal.  Lens-source
pairs are assigned weights according to the error on the shape
measurement via
\beq
w_{ls} = \frac{\Sigma_c^{-2}}{\sigma_s^2 + \sigma_{SN}^2}
\eeq
where $\sigma_{SN}^2$, the intrinsic shape noise, was determined as a
function of magnitude in M05, figure 3.  The factor of
$\Sigma_c^{-2}$ downweights pairs that are close in redshift.

Once we have computed these weights, we compute the lensing signal in
53 logarithmic radial bins from 20 $h^{-1}$kpc to 4 $h^{-1}$Mpc as
a summation over lens-source pairs via:
\beq
\ds(R) = \frac{\sum_{ls} w_{ls} \gamma_t^{(ls)} \Sigma_c}{2 {\cal
    R}\sum_{ls} w_{ls}} 
\eeq
where the factor of 2 arises due to our definition of ellipticity.

There are several additional procedures that must be done when
computing the signal (for more detail, see M05).  First, the signal
computed around random points must be subtracted from the signal
around real lenses to eliminate contributions from systematic shear.
Second, the signal must be boosted, i.e. multiplied by $B(R) =
n(R)/n_{rand}(R)$, the ratio of the number density of sources relative
to the number around random points, in order to account for dilution
by sources that are physically associated with lenses, and therefore
not lensed.

In order to determine errors on the lensing signal, we divide the
survey area into 200 bootstrap subregions, and generate 2500
bootstrap-resampled datasets.  

We also would like to account for the total calibration uncertainty of
roughly 8 per cent at the $1\sigma$ level (including redshift
uncertainties, shear calibration, stellar contamination, and all other
effects that cause calibration bias).  In principle, this can be done
by multiplying the signal in each bootstrap resampled dataset by
$1+0.08g$ where $g$ is a Gaussian random number with mean zero and
variance $1$, and the net effect will be to cause correlations of
order 8 per cent between radial bins in addition to any correlations
due to shape noise, systematic shear, or other causes.  However, we
find that this procedure causes a significant bias in our fits,
because points that are naturally below the best-fit signal have their
errorbars increased by a smaller amount than those that are above the
best-fit signal, and therefore those that are below the real signal
are weighted more highly in the fits.  Consequently, we generate the
bootstrap-resampled datasets without including the calibration
uncertainty, and add the 8 per cent uncertainty in quadrature with the
statistical error on the mass determination.

Furthermore, to decrease noise in the covariance matrices due to the
bootstrap, we rebin the signal into 18 radial bins total (of which 12
are in the range of radii that we will eventually elect to use for the
fits).

\subsection{Theoretical systematics}\label{SS:sys}

There are several theoretical systematics that are particularly problematic on
small scales near massive galaxies.  Here we describe two of them,
estimate their magnitude, and explain how we correct for them.

\subsubsection{Non-weak shear}

\begin{figure}
\includegraphics[angle=0,width=3.2in]{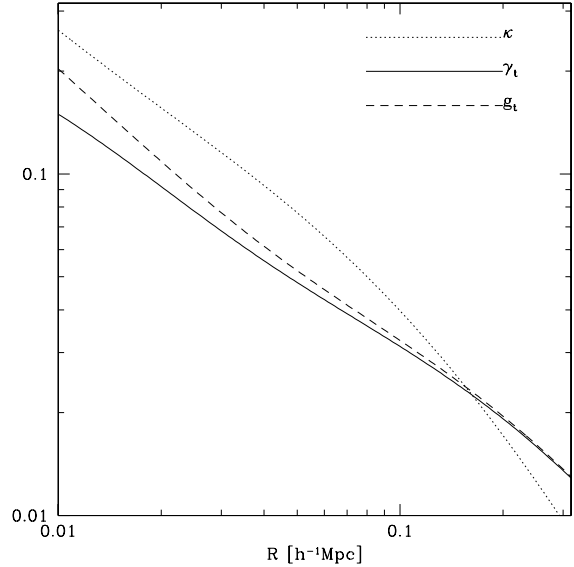}
\caption{\label{F:params}$\kappa$, $\gamma_t$, and $g_t$ for the
  typical density profile described in the text, as labelled on the
  plot.} 
\end{figure}

The first small scale systematic we consider is the fact that for
these relatively massive halos, the assumption of weak shear may no
longer hold on the smallest scales that we probe (20 \hkpc).  In
reality, what is measured in g-g lensing measurements is the reduced
shear, 
\begin{equation}
g_t = \frac{\gamma_t}{1-\kappa},
\end{equation}
(where $\kappa=\Sigma/\Sigma_c$). Hence, $g_t \approx \gamma_t$ to a high
degree of accuracy in the $\gamma_t
\ll 1$, $\kappa \ll 1$ regime 
that is typically probed with weak lensing.   Fig.~\ref{F:params} includes
a plot of $\gamma_t$, $\kappa$, and $g_t$ for a typical LRG profile
with the same density profile parameters as in Fig.~\ref{F:rho},
without adiabatic 
contraction (AC makes no perceptible difference in these quantities
on these scales).  As shown there, at the minimum scale at which we
attempt to measure the lensing signal (20 $h^{-1}$kpc) the
difference between $\gamma$ and $g$ is $\sim 15$ per cent, and at 40
$h^{-1}$kpc it is 9 per cent, causing us to
overestimate \ds.  

Unfortunately, it is difficult to take into account the effect of
non-weak shear on our shear estimator, $\langle e_+\rangle$, in a way
that is completely model-independent.  Appendix~\ref{A:nonweak} shows
the rather involved calculations necessary to estimate the importance
of any terms beyond first-order in $\gamma_t$ and $\kappa$.  As we
show there (Eq.~\ref{E:nonweak}, reproduced here for convenience), the
second-order calculation of the value of the shear estimator is
\beq
\langle e_+ \rangle = 2{\cal R} \langle\Delta\Sigma\rangle
\langle\Sigma_c^{-1}\rangle \left( 1 +
  \frac{\langle\Delta\Sigma\,\Sigma\rangle}
  {\langle\Delta\Sigma\rangle}\frac{\langle \Sigma_c^{-2} \rangle}
  {\Sigma_c^{-1}}\right),
\eeq
which gives a correction larger than the naive $1+\kappa$.

The two fractions in the correction term need to be estimated in some
manner.  The first one can be estimated using our earlier model of
averaging $\Delta\Sigma$ over the halo mass function from the relevant
minimum mass to the maximum mass, and is found to be $\approx 1$ for
the faint LRG sample, and $\approx 1.03$ for the bright LRG sample.  The
second fraction in the correction term,
$\langle\Sigma_c^{-2}\rangle/\langle\Sigma_c^{-1}\rangle$, can be very
simply calculated when the signal is calculated, where the averages
are computed using the weighting used to get the lensing signal, and
typically equals approximately 1.05.
Then, when the fitting is completed, we multiply the model
$\Delta\Sigma$ by the correction term. 

\subsubsection{Magnification bias}

The next issue we consider is the
effect of magnification bias on the boost factors.  Magnification
bias will tend to increase or decrease the number density of observed
sources relative to that expected from the random catalogs; hence, our
procedure of multiplying the signal by $n(R)/n_{rand}(R)$ in order to
correct for dilution by physically associated (non-lensed) sources will tend to
overestimate the signal, since some of those additional sources are
actually lensed.  The degree of this overestimation was computed for
each source sample as a function of its apparent magnitude and
resolution factor
distributions, and
was found in M05 to be $\delta n/n = 1.9\kappa, 0.7\kappa, 3.1\kappa$ for
$r<21$, $r>21$, and LRG sources respectively.  (Note that $\langle
\Sigma_c^{-1}\rangle$ varies with source sample as well, which gives
different $\kappa$ for each sample.)  In order to correct for this
overestimation, we multiply the signal for each model by
$(1+a\Sigma\langle \Sigma_c^{-1}\rangle)$ before comparing against the
real data, where $a$ and $\langle\Sigma_c^{-1}\rangle$ are found as a
weighted average over the values for each source sample.
Fig.~\ref{F:params} shows, for the model density profile used
throughout this section, the value of $\kappa$ from which the size of
this correction factor can be ascertained.

A related issue is that of the effect of
magnification bias on the source redshift distribution.  This effect
is not a problem for the $r<21$ or the LRG source sample, since we use
photometric redshifts that should still be reliable,
so it is only a concern for the $r>21$ sample for which we use a
distribution $p(z)$.  Since
magnification bias allows the observation of sources that would not
have otherwise been seen, it means that in regions where magnification
bias is significant, we tend to underestimate the source redshifts,
and therefore $\Sigma_c^{-1}$, overestimating the signal \ds.  The
mean redshift of the $r>21$ sample is $\langle z\rangle=0.45$. 

For the model LRG signal used in this section, we estimate the size of
this effect as follows.  We consider a moderately large value of
$\kappa=0.1$ such as may be present for transverse separations of 40
$h^{-1}$kpc (see Fig.~\ref{F:params}).  For the $r>21$ sample, this
means that $\delta n/n ~ 0.07$, so the fraction of sources used that
have been scattered into the sample is $0.07/1.07 \sim 0.065$.  
We consider what happens if those sources are actually all at $z=1$
(more than twice the mean redshift of this sample, giving a
conservative estimate).  $\Sigma_c^{-1}$ for these sources is then
actually 50 per cent larger than what we have assumed.  If 6.5 per
cent of the sample has had its $\Sigma_c^{-1}$ value so
underestimated, this means that the sample average $\Sigma_c^{-1}$
value is thus 3 per cent too small, and the signal has been
overestimated by this amount.  Since this value is far smaller than
other sources of systematic and statistical error, and is a
conservative estimate, we do not apply a
correction.

\section{Results}\label{S:results}

Those who are uninterested in the details of the lensing systematics tests
in the following subsections may wish to skip directly to the results
of the fitting, in subsection~\ref{SS:fits}.  

\subsection{Small-scale software-related systematics}\label{SS:resultsys}

One systematic potentially affecting all source samples on small
scales is the sky subtraction systematic, an error in the sky
estimation near bright lenses
\citep{2005MNRAS.361.1287M,2006ApJS..162...38A}.  A similar 
systematic is the possibility of errors in deblending.  To test for
these problems, we use simulations 
that were used for the small-scale LRG autocorrelation function
measurement in \cite{2005astro.ph.12166M}.  These simulations were
created by placing fake LRGs 
on real SDSS images in run 2662, camcol 1.  For that paper, the issue
of concern was the effect on a measured LRG
flux due to the presence of another nearby LRG, so two fake LRGs were
placed in each field at various 
angular separations.  For this work, similar simulations were created
with only one fake LRG per SDSS field, with 470 fields total.  The
simulated LRG flux is $r=17.9$, 
 corresponding to a $M_r=-22.5$ LRG at $z=0.3$ with $r_{deV}=3.4$''.
 The axis ratio $b/a=0.7$, with quantized random orientation relative
 to the scan direction (i.e., it was allowed to have any of 360
 orientations evenly spaced in position angle).  We used these
 values of flux and size to place conservative constraints on
 software-induced systematics; studying the variation with LRG size
 and flux is unfortunately 
 prohibitive due to computer time and storage space limitations. 

The new
images were then processed with the same version of the SDSS
processing software, {\sc Photo}, as the original data (rerun 137).
We use these results to decide 
which ranges of transverse separations to use in the real data.  The
 results presented in this paper are averaged over five separate
 ``reruns'' with different simulated LRG positions but the same
 photometric properties.
%

There are several concerns for which we would like to test:
\begin{enumerate}
\item Over- or under-estimation of the sky can lead to modulation of
  the number density near LRGs (because they change the flux and
  apparent size, and hence scatter galaxies into or out of our
  catalog).  Since we assume that any change in 
  number density relative to that around random points is due to
  physically-associated sources, this problem may cause a
  bias in the lensing signal.
\item If the deblender or sky determination algorithm includes a
  spurious gradient in the flux around LRGs, it can lead to the
  galaxies near the LRG (whether physically-associated or not) having
  some spurious additive and/or multiplicative shear.
\item If the LRGs cause nearby galaxies to have mis-measured fluxes,
  then depending on the way this effect varies with band, it may
  change the colors and hence lead to inaccurate photometric redshift
  determination.  This problem would affect the $r<21$ and LRG source
  samples.  Furthermore, for the $r>21$ sample, for which we assume
  $dn/dz$, the redshifts of the sources may be underestimated, and
  hence $\Delta\Sigma$ overestimated, by some amount.
\end{enumerate}

\begin{figure*}
\includegraphics[angle=0,width=5.5in]{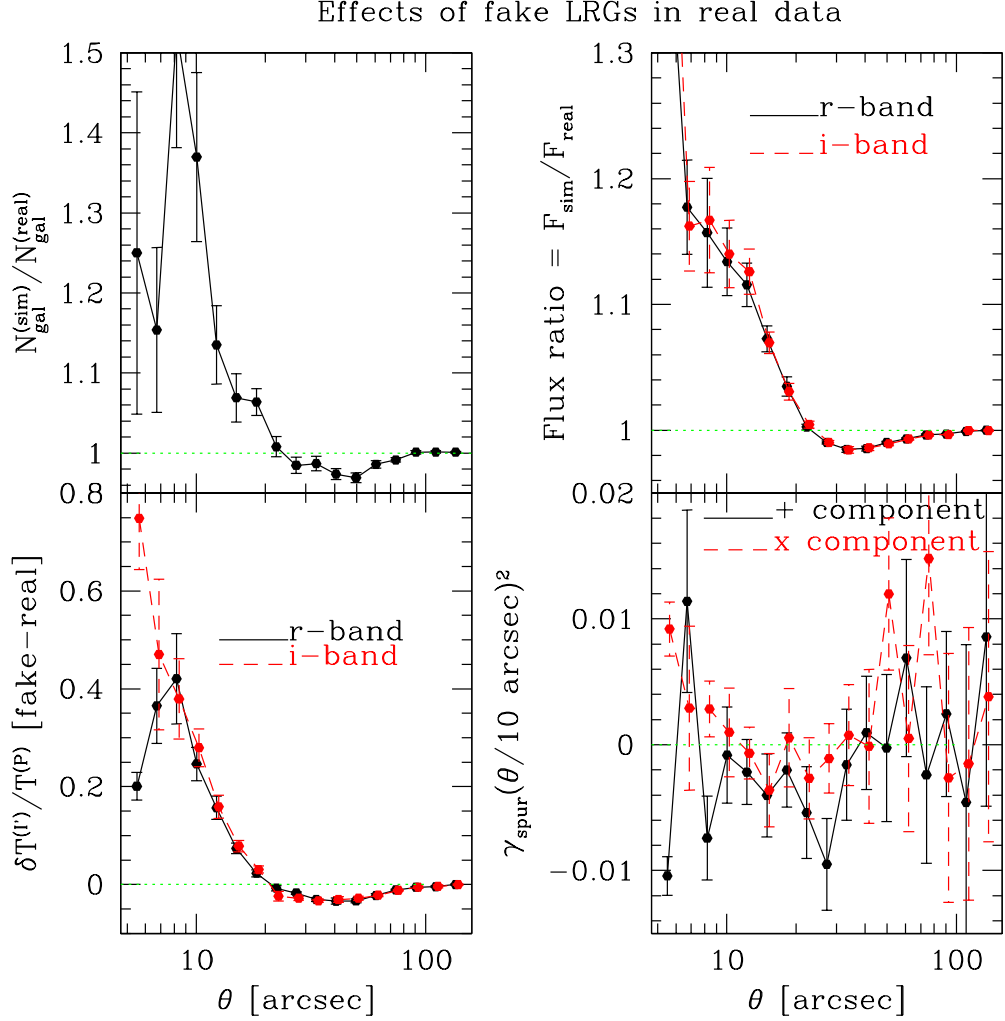}
\caption{\label{F:simall}Four plots showing the effect of simulated
  LRGs on real galaxies in SDSS data.  In all plots, the dotted line
  shows the nominal value  if the
  simulated LRGs had no effects on the photometry of the real
  galaxies.  The upper left panel shows the effect on the number
  density of real galaxies in our lensing source catalog around the
  positions of fake LRGs.  The upper right panel shows the effect of
  the simulated LRG on the real galaxy fluxes in both $r$ and $i$
  bands as labelled.  The lower left panel shows the effect on the
  apparent size of the real galaxies relative to the size of the PSF,
  where the quantities measured are described in more detail in the
  text.  Finally, the lower right panel shows the spurious shear
  caused by the deblender or the sky subtraction errors; in order to
  make the results easier to view, this quantity has been multiplied
  by $(\theta/10'')^2$.}  
\end{figure*}

In order to test for these effects, we searched for real galaxies
  around the positions of the simulated LRGs in the real data and the
  simulations.  The first test, for modulation of number density, was
  performed simply by comparing the number of real galaxies around
  those locations in the simulation versus in the real data; results
  are shown in the upper left panel of Fig.~\ref{F:simall}. 
As shown there, there is the same depletion in number density on
30--90$''$ scales as noted in previous works, with a crossover point
occuring between 20--30$''$, and an excess of galaxies below that point.
These findings are consistent with the sky being underestimated (and
hence source galaxy fluxes overestimated) below the crossover point,
and overestimated above that point.  
We note that the crossover point around $23''$ corresponds
to $\sim 93$ comoving $h^{-1}$kpc at the sample mean redshift.  Also,
in this panel as for the others, the results in the simulations match
those in the real data above $100''$.

Next, we calculated both the tangential and 45 degree shear components around the
positions of the fake LRGs, using only those real galaxies that passed
cuts to be in our source catalog in both the real data and the
simulation.  The importance of this cut is that if we use the same real galaxies
in both samples, then when we take the difference between the shears
in these samples, all shape noise will be eliminated, giving an
estimate of the ``systematic'' shear in both components.  The resulting plot is
shown in the lower right panel of Fig.~\ref{F:simall}.
The $x$ shear component, while containing several points that are
slightly discrepant from zero, is nonetheless consistent with zero
spurious shear for all points above 10''; below this scale, the
errorbars on both shear components appear to be underestimated.  The
$+$ shear component, 
the tangential shear, is also consistent with zero down to 10'', or
roughly 40 $h^{-1}$kpc: the constraint $|\gamma_{sys}|<0.004$
corresponds to $|\Delta\Sigma_{sys}| < 20$, which is far smaller than
the statistical errors. Below this scale,  constraints on contamination are
weaker, so we use a minimum transverse separation of $40$ \hkpc\ for
all fits.

This calculation only shows additive shear on galaxies that
are matched.  In the real
data, any additive shear determined in this way will have to be
multiplied by the boost factor $B(R)$, since physically-associated
sources are equally susceptible to software-induced systematics.
Furthermore, this calculation only applies to galaxies that are
matched in both the real and simulated data; it does not apply to
galaxies that are scattered into the catalog due to sky
underestimation.  It is plausible that there is some
selection bias operating in the selection as these galaxies, for
instance causing us to select those that are aligned tangentially to
the lens rather than radially, that would cause an additional bias in
the results.  Unfortunately, placing constraints on these galaxies
is much more demanding because there are no matching galaxies in the
real data with which they can be paired to eliminate shape noise.
Finally, this calculation tells us about additive shear errors, but
not about multiplicative biases due to the deblender. 

The next question we address is that of the effect on the fluxes.
While it is clear that either the fluxes or apparent sizes of sources
near the fake LRGs is affected (since the number density of sources
changes near them), the change in the fluxes may also affect
classification into our three source samples, and may change
photometric redshifts for the two samples that use them and the source
redshift distribution for the $r>21$ sample.  For the galaxies that
match in the real versus simulated data, the upper right panel of
Fig.~\ref{F:simall} shows the ratio 
of measured flux in the simulated data versus in the real data, in
both $r$ and $i$ bands.  
As shown, at 10'' ($\approx 40 h^{-1}$kpc), the fluxes are
overestimated by roughly 15 per cent, corresponding to a change in
apparent magnitude of 0.15.  Interestingly enough, the change is
 consistent for the $r$ and $i$ bands, even though the
simulated LRG fluxes in these bands are not the same.  This suggests
that the effect 
may not noticably change photometric redshifts since they depend on
color rather than magnitude.

Finally, while we may guess that the change in number density is only due
to the shift in the apparent magnitudes, the lower left panel of
Fig.~\ref{F:simall} indicates 
that the apparent sizes of the galaxies relative to the sizes of the
PSF are also changed in a way that can cause the number densities to
change as in the upper left panel.
The quantity shown in this plot is
\beq
\frac{T^{(I)}_{sim} - T^{(I)}_{real}}{T^{(P)}} = \frac{\delta T^{(I)}}{T^{(P)}},
\eeq
or the change in the trace of the adaptive moment matrix due to
presence of the LRG, relativev to the trace of the adaptive moment
matrix of the PSF.  
When we compare against Eq.~\ref{E:R2def}, we see that the change in
resolution factor $R_2$ due to the change in apparent size is
\beq
\delta R_2 = \frac{\delta T^{(I)}}{T^{(P)}}(1-R_2)^2
\eeq
So, for a typical galaxy with $R_2=0.55$ (sample median) at 10'' from
one of these LRGs, the resolution factor is increased by 0.09 to 0.64.
A typical galaxy at separations of 60'' has resolution factor
decreased by -0.01 to 0.54. 

Finally, we note for the sake of completeness that this simulation was
only for one type of LRG, $r=17.9$, with $r_{deV}=3.3$''. This is
reasonably close to the mean apparent magnitude of the spectroscopic
LRGs used for this analysis. We have not
derived variations of the effect with total LRG flux 
or apparent size, nor have we done any calculations for galaxies with
profiles that are not purely de Vaucouleurs, so these results should
not be used for, e.g., the Main galaxy sample which has a higher flux
on average, and is a mixture of both de Vaucouleurs and exponential
profile galaxies. 

As a result of the findings in this section that deblending and sky
subtraction systematics become significantly worse in a way that is
difficult to quantify below about 10'' (or 40 \hkpc), in the analysis
that follows, we only use the lensing signal above 40 \hkpc\ for
fits even though plots show signal down to 20 \hkpc\ .  We correct
the signal for the overestimation of 
the boosts due to the modulation of the number density near the LRGs,
but because this correction is so large and uncertain below 10'', we
only show the results below that separation to demonstrate consistency
with the results derived from higher angular separations.  We also
increase the error bars to account for systematics, with an increase
of 50 per cent at 20 \hkpc\ down to no increase at 50 \hkpc.

\subsection{Signal amplitude}
\label{ss:amplitude}

In this section we discuss the lensing signal calibration for each
source sample.  The standard systematics tests 
as described in \cite{2005MNRAS.361.1287M} were performed to ensure
that systematics are under control (45-degree test, random catalog
test), and no anomalies were found.

One test that we focus on in particular here is the ratio test, which
allows us to compare the effective average amplitude of $\Delta\Sigma$
for different source samples.  The ratio test is of particular
interest because, for \cite{2005MNRAS.361.1287M}, the lensing signal
for the three source samples was found to be consistent at the
$1\sigma$ level, but  with lenses at a median redshift of
$0.1$.  For this work, the lens redshift range is from $0.15$ to
$0.35$, with an effective mean of $0.24$.  The peak redshift of the
$r<21$ sample is $0.35$, quite
close to the lenses, which means that for this work, we are far more
sensitive to the details of the photometric redshift error
distribution than for \cite{2005MNRAS.361.1287M}.  In contrast, the
mean redshift of the $r>21$ sample is $0.45$ and of the LRG source
sample is $0.55$, so we expect that the signal with these two source
samples will still be fairly reliable.

\begin{figure}
\includegraphics[angle=0,width=3.2in]{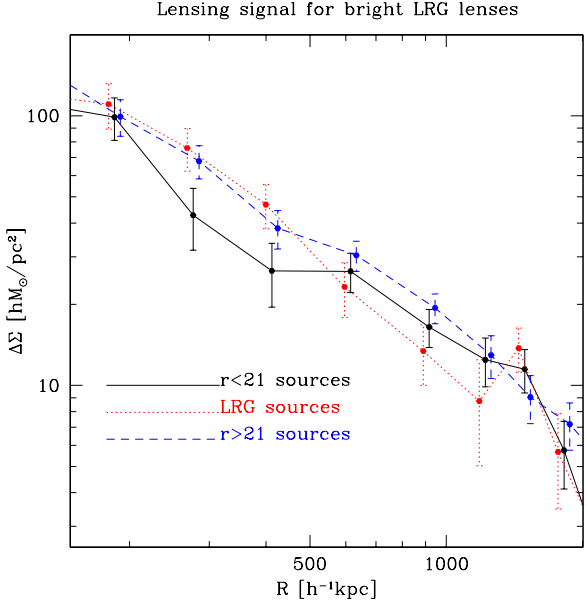}
\caption{\label{F:lrgdiffsource}Lensing signal $\Delta\Sigma$ for the
  bright lens sample, with three different source samples as indicated
  on the plot.}  
\end{figure}
Fig.~\ref{F:lrgdiffsource} shows the lensing signal for the bright LRG lens
sample with the $r<21$, $r>21$, and LRG source samples separately for
$150 < R < 2000$ \hkpc.  We use these scales because they are the most
free of systematics and offer the highest signal-to-noise.  It
seems clear that the LRG and $r>21$ samples effectively agree in
average amplitude, but the $r<21$ sample gives a lower overall calibration.
When we determine an average effective amplitude of the lensing signal
for each lens sample (``faint,'' ``bright,'' and ``all'') for each
bootstrap dataset, and compare this amplitude 
for the different source samples, we find the following
relations hold:
\begin{align}\label{E:calib}
\frac{\langle\Delta\Sigma\rangle_{r>21}}{\langle\Delta\Sigma\rangle_{LRG}}
&= 0.98\pm 0.11, &
\frac{\langle\Delta\Sigma\rangle_{r<21}}{\langle\Delta\Sigma\rangle_{r>21}}
&= 0.84\pm 0.11\mbox{ (faint)} \\ 
\frac{\langle\Delta\Sigma\rangle_{r>21}}{\langle\Delta\Sigma\rangle_{LRG}}
&= 1.01\pm 0.09, &
\frac{\langle\Delta\Sigma\rangle_{r<21}}{\langle\Delta\Sigma\rangle_{r>21}}
&= 0.86\pm 0.08\mbox{ (bright)} \\ 
\frac{\langle\Delta\Sigma\rangle_{r>21}}{\langle\Delta\Sigma\rangle_{LRG}}
&= 1.00\pm 0.07, &
\frac{\langle\Delta\Sigma\rangle_{r<21}}{\langle\Delta\Sigma\rangle_{r>21}}
&= 0.85\pm 0.06\mbox{ (all)} 
\end{align}

As shown in Eq.~\ref{E:calib}, the signal for the LRG and $r>21$
source samples is statistically consistent for both LRG lens samples,
but in contrast, the signal with the $r<21$ (photoz) sources is about
15 per cent lower, a $2.5\sigma$ discrepancy when considering all
lenses together.  This finding suggests that for the $r<21$ sample,
some problem with the photometric redshift error distributions is
causing $\Sigma_c^{-1}$ to be overestimated by 15 per cent, thus
underestimating $\Delta\Sigma$ by that amount. 

In order to test this hypothesis, we further split the $r<21$ sources
into two samples: those with photometric redshift lower than $0.4$,
and those with photometric redshift higher than $0.4$.  When we
compare the signal with the ``low redshift'' and ``high redshift''
$r<21$ sources, we find the following relations:
\begin{equation}\label{E:photozhighlow}
\frac{\langle\Delta\Sigma\rangle_{z_s>0.4}}{\langle\Delta\Sigma\rangle_{z_s<0.4}}=
\begin{cases}
0.71\pm 0.13, &\text{Faint LRG lenses;} \\
0.80\pm 0.12, &\text{Bright LRG lenses;} \\
0.76\pm 0.09, &\text{All LRG lenses.}
\end{cases}
\end{equation}
The net result, then, is that the sources with photometric redshift
larger than $0.4$ give signal that is 24 per cent lower than the signal
with photometric redshift lower than $0.4$.  When we take into account
that this higher photometric redshift sample has more weight than the
lower photometric redshift sample, this explains the signal with the
full photometric redshift sample being low by 15 per cent.  This
result also
explains our failure to notice this discrepancy in previous works; for
\cite{2005MNRAS.361.1287M} and \cite{2005astro.ph.11164M}, the mean lens
redshift was lower, so (a) the sample 
with photometric redshift lower than $0.4$ had significantly higher
weight than the sample with photometric redshift greater than $0.4$;
and (b) an imperfect understanding of the photometric redshift
bias and scatter at photometric redshift larger than $0.4$ gives
a smaller fractional error on $\Sigma_c^{-1}$ for lower lens
redshift.  Consequently, the signal with the $r<21$ sample was not as
noticably discrepant in previous works.

When considering $\Sigma_c^{-1}(z_l=z_{eff},z_s)$ as a function of
source redshift, we note that for galaxies at $r<21$,
a net bias of 0.04 would be sufficient to achieve a 15 per cent change
in overall calibration, e.g. if we assume that galaxies really at
$z=0.4$ are at $z=0.44$.  Alternatively, a smaller bias of $\sim 0.02$
and a modest
increase in the net scatter of $0.02$ will tend to cause the same change in
calibration.  Since the peak of the redshift distribution
for these galaxies is around $z=0.35$, the photometric redshift error
distribution was determined with a large number of galaxies from
$z=0.3$--$0.4$, but with a much smaller number of galaxies at $z>0.4$, and
hence a bias and scatter increase of these magnitudes is within the
statistical error.

Considering that we can fairly easily explain this bias in the $r<21$
sample relative to the more reliable (for this purpose) $r>21$ and LRG
source samples, before doing the fits we have increased the signal
with the $r<21$ sample by 15 per cent as in Eq.~\ref{E:calib}, and
increased the statistical error by 6 per cent to account for the
statistical error in the determination of this calibration factor. In
practice, the increase in statistical error is achieved by multiplying
the signal in individual bootstrap samples by a random  number
of mean $1$ and Gaussian standard deviation $0.06$.  Note that this calibration
change affects the mass determination, not the profile shape.

\subsection{Intrinsic alignments}\label{SS:iaresults}

There are two intrinsic alignment effects that are important for
lensing, but only one is important for galaxy-galaxy lensing as
presented here.  The intrinsic
ellipticity-intrinsic ellipticity (II) correlation is
not important for galaxy-galaxy lensing, because we average over all
relative lens-source ellipticity orientations.  The second effect,
which is important for this work, is an alignment between the
intrinsic ellipticity and the local density field, or the GI
correlation.  This effect comes into 
play because we necessarily include some physically-associated pairs
(i.e., pairs of lenses and ``sources'' that are really part of the same
local structure), so if those sources
have a tendency to align tangentially or radially relative to the
lens, they will provide an additive bias to the lensing signal.

Correlations between intrinsic ellipticity and density have been
demonstrated robustly using bright, red galaxies on large scales
\citep{2006MNRAS.367..611M} and on small scales
\citep{2005astro.ph..7108M,2006astro.ph..3471D,2006MNRAS.369.1293Y},
with the effect being more controversial for the general galaxy
population and for spirals.  In the cases
considered in these papers, the typical scenario is that a large,
bright galaxy is found to point preferentially in the 
direction of nearby overdensities, e.g., along a cluster major axis.
In our case, we are concerned that the fainter sources around a
large, bright galaxy may be pointing preferentially radially or
tangentially towards it; unfortunately,
constraints on this scenario are weak and/or difficult to convert to
measurements of lensing shear
\citep{2001ApJ...555..106L,2002AJ....124..733B,2004MNRAS.353..529H}.
We expect that on scales larger than $\sim 100$ \hkpc, 
above which the boost factors are fairly small, there will be little
contribution from intrinsic alignments.

Here, we try to measure the intrinsic alignment contamination of the
lensing signal using photometric redshifts.  For a given lens, the
sources with $r<21$ are split into a ``physically associated'' (PA)
sample with $z_l-0.1 < z_s < z_l+0.1$ and a ``lensed'' (L) sample with
$z_s \ge z_l+0.1$.  The difficulty, of course, is that the photometric
redshifts have a broad error distribution, so that both of these
samples are going to have some galaxies in them that do not actually
fit the criteria of being PA or L; in the two samples, the fractions
of physically associated sources are $f_{PA}^{(PA)} =
(B^{(PA)}-1)/B^{(PA)}$ and $f_{PA}^{(L)} = (B^{(L)}-1)/B^{(L)}$.  We
measure signals $\hat{\gamma}^{(PA)}$ and $\hat{\gamma}^{(L)} =
\Delta\Sigma^{(L)}\langle\Sigma_c^{-1}\rangle^{(L)}$ that are related
to the ``real'' lensing and PA shears via
\begin{align}
\hat{\gamma}^{(PA)} &= f_{PA}^{(PA)}\gamma_{PA}^{(PA)} +
(1-f_{PA}^{(PA)})\gamma_L^{(PA)} \\ \notag
 &= f_{PA}^{(PA)}\gamma_{PA}^{(PA)} +
 (1-f_{PA}^{(PA)})\Delta\Sigma_L^{(L)}\langle\Sigma_c^{-1}\rangle^{(PA)} \label{E:gammahatpa}
\end{align}
where the lensing shear in the PA sample can be determined using
$\Delta\Sigma$ from the lensing signal and multiplying by the average
$\Sigma_c^{-1}$ for the PA sample.  This average value can be
determined by integrating over the known lens redshift distribution,
source photometric redshift distribution, and photometric redshift
error distribution, and is found to be a factor of $2.5$ lower than
$\langle\Sigma_c^{-1}\rangle^{(L)}$, partly due to the
significant number of sources in the PA sample that are foregrounds
wth $\Sigma_c^{-1}=0$.  Thus, we can rearrange this
equation to get $\gamma_{PA}^{(PA)}$, the intrinsic alignment shear in
the PA sample, from
the measured shear with the physically assocated sample.

Next, we can estimate the contamination to the real lensing signal
using the following equation:
\begin{align}
\hat{\gamma}^{(L)} &= B^{(L)}\left[(1-f_{PA}^{(L)})\gamma_{L}^{(L)} +
f_{PA}^{(L)}\gamma_{PA}^{(L)}\right] \\ 
 &= \gamma_{L}^{(L)} +
\frac{f_{PA}^{(L)}}{1-f_{PA}^{(L)}}\gamma_{PA}^{(L)} \label{E:gammahatle}
\end{align}
We note several things about this equation. The first term is simply
the lensing signal, where the boost factor is necessary to account for
the fact that not all the sources are lensed.  The second term is the
intrinsic alignment contamination term.  We have pointed out with our
notation a nuance
that may be important: in Eq.~\ref{E:gammahatpa}, we will be able
to determine the intrinsic alignment shear due to physically
associated sources in the PA sample; in Eq.~\ref{E:gammahatle}, we
actually need the intrinsic alignment shear due to physically
associated sources in the L sample.  We return to this difference shortly.

\begin{figure}
\includegraphics[angle=0,width=3.2in]{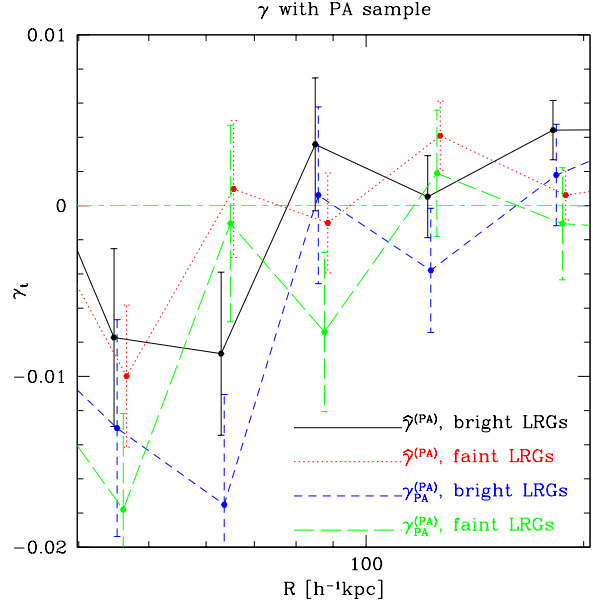}
\caption{\label{F:allia}Measured shear with the PA source sample for
  both LRG lens samples, and the implied shear due to intrinsic
  alignments after subtracting off the contribution due to lensing.} 
\end{figure}
Fig.~\ref{F:allia} shows the measured shear with the PA samples,
$\hat{\gamma}^{(PA)}$ for the two lens samples, and then the corrected
$\gamma_{PA}^{(PA)}$ after we remove the contamination due to
lensing.  As shown, the signals are consistent with zero
above 100 \hkpc; below that scale, they are consistently negative,
indicating a tendency for satellites to ``point'' radially towards
primaries.  The $\chi^2$ for a fit to zero signal for the two lens
samples are $11.6$ and $12.5$ for the bright and faint lens samples
respectively (with 3 degrees of freedom with $40<R<100$ \hkpc), giving
a probability to 
exceed by chance of 0.01 and 0.007, indicating a positive detection on
these scales.  If we average the value of
$\gamma_{PA}^{(PA)}$ for those scales, we get averages
of $-0.0084\pm 0.0034$ and $-0.0086\pm 0.0030$.  We can also
get the projected lensing 
contamination, if we make the crucial assumption that
$\gamma_{PA}^{(PA)} = \gamma_{PA}^{(L)}$,  by using
$\langle\Sigma_c^{-1}\rangle$ for the  $r<21$ sample of $1.9\times
10^{-4}$.  This signal, when 
subtracted from the measured $\hat{\gamma}^{(L)}$ to get a
  decontaminated lensing signal, leads to an increase of the signal
  roughly equivalent to the size of the $1\sigma$ errors or larger on
  these scales, and hence has a significant effect on fits to the
  profile.

In comparison with recent works, we note that using SDSS data,
\cite{2005astro.ph..9405A} also found a tendency for satellites to
align radially towards primaries using spectroscopically-determined
primary-satellite pairs.  Over the range $7 < R < 50$ \hkpc,
they found a mean tangential shear of $\gamma_t = -0.030\pm 0.007$.
This number is, as expected, larger than our measured tangential
shear, since ours is at a mean transverse separation that is close to
three times as high as the mean separation for their measurement.  

However, we must consider further whether this is really the proper
correction to apply.  For the $r<21$ sample, the sources that are
truly physically associated in the L sample are, by definition, those
with poor photometric redshifts.  Hence, studies of the photometric
redshifts that we use in comparison with DEEP2 redshifts
\citep{2005MNRAS.361.1287M} indicate that these will tend to be, on 
average, bluer and fainter.  On the other hand, the physically
associated sources in the PA sample are those with good photometric
redshifts, which will tend to be brighter and redder.  There is no
{\it a priori} reason to suppose, then, that
$\gamma_{PA}^{(PA)}=\gamma_{PA}^{(L)}$, in light of the fact that many
intrinsic alignment measurements indicate that the effect depends
sensitively on the galaxy luminosity and color.  In fact, both large-
and small-scale
measurements seem to indicate that intrinsic alignments are negligible
for bluer, fainter galaxies and large for bright, red ones, which
suggests that $\gamma_{PA}^{(L)}$ may well be zero, with no correction
necessary \citep{2006astro.ph..7139A,2006MNRAS.367..611M,2006MNRAS.369.1293Y}. 

The bottom line, then, is that it is not clear how to correct the
signal with $r<21$ sources for intrinsic alignment contamination in a
way that is not based on many tenuous assumptions.  We may come to a
similar conclusion about the $r>21$ sample.  Since we only have a
redshift distribution for this sample, the sample
of galaxies that are physically associated are a mix of blue and red,
but are on average (at a given redshift) fainter than the physically
associated sources from the $r<21$ sample, and hence the corrections
derived from the $r<21$ PA sample may not apply.

Because of this problem with both the $r<21$ and the $r>21$ samples with 
respect to intrinsic alignments on $R<100$ \hkpc\ scales, we choose to 
rely solely on the high-redshift LRG sources on these scales.  While the 
high-redshift LRG sample contains, by definition, bright red galaxies that 
may be expected to have strong intrinsic alignments, they are located at 
redshifts $z>0.4$, whereas our lenses are in the range $0.15<z<0.35$.  
Furthermore, we impose a cut on the source photometric redshift ($z_s > 
z_l+0.1$) to further eliminate low-redshift contamination.  The net effect 
is that the physically associated fraction at $R>40$ \hkpc\ is negligible 
when using this source sample, so it must be free of intrinsic alignments.  
To ensure that the signal on $40$--$100$ \hkpc\ scales, which may be 
crucial to understanding the profile, is as free of systematics as 
possible (while retaining higher statistical power on larger scales), we 
thus use {\it only} the LRG sources at $R<100$ \hkpc, and average all the 
source samples for $R>100$ \hkpc.

\subsection{Fits to lensing signal}\label{SS:fits}

We begin the fits with the simplest possible way, and add parameters
until the best-fit $\chi^2$ does not decrease with the addition of
parameters.  Table~\ref{T:fitresults} shows the results for best-fit
parameters for the fits as enumerated and described there. In this
table, in all cases the fits are from a minimum scale of $40$ \hkpc\
(to avoid systematics-dominated inner regions) to a maximum of $2$
\hmpc\ (to avoid the need for complicated modeling of the halo-halo
term).   Errors on all parameters are the
formal $1\sigma$ errors 
from the fit covariance matrix determined using derivatives of the
$\chi^2$ with respect to each fit parameter.  We remind the reader
that the mass estimates have had an additional 8 per cent systematic
error added in quadrature with the formal statistical error, and the
fits are performed using the full covariance matrix (which includes
minimal correlations at small scales, and correlations as large as 20--30
per cent at large scales).
\begin{table*}
\caption{\label{T:fitresults}Fits performed, including identifying
  number, and fit parameters.  Bold-faced fit parameters are meant to
  indicate that they were held fixed, whereas those in regular font
  are those that were allowed to vary, with $1\sigma$ errors.  For all fits,
  $R_{min}=0.04$ \hmpc. Note that the fitted values for concentration 
should be increased by about 20 per cent  
when compared to simulations to account for the scatter in cluster profiles, 
which was found to reduce the value in the lensing 
fits. Similarly, the fitted masses correspond
better to the median rather than mean mass \protect\citep{2005MNRAS.362.1451M}.  }
\begin{tabular}{rrcccccccc}
\hline\hline
Fit & Description & $\alpha$ & $\beta$ & $c$ & $M$ & $M_{stellar}$ &
$R_{s}$ & $\chi^2$ & dof  \\
 & & & & & $10^{13}h^{-1}M_{\odot}$ & $10^{11}h^{-1}M_{\odot}$ &
 $\!\!$\hkpc$\!\!$ & & \\
\hline \\
\hline
\multicolumn{10}{c}{Faint LRG sample} \\
1 & SIS & $\mathbf{-2.0}$ & $\mathbf{2.0}$ & $\mathbf{10^4}$ & $2.4\pm
0.3$ & $\mathbf{2.2}$ & $\mathbf{8}$ & $18.6$ & $10$ \\
2 & Power-law & $\mathbf{-2.0}$ & $1.85\pm 0.06$ & 
 $\mathbf{10^4}$ & $3.0\pm 0.4$ & $\mathbf{2.2}$ & $\mathbf{8}$ &
 $11.5$ & $10$ \\
3 & NFW & $\mathbf{-1.0}$ & $\mathbf{3.0}$ & $4.0\pm 0.6$ & $2.9\pm 0.4$ & 
   $\mathbf{2.2}$ & $\mathbf{8}$ & $11.5$ & $10$ \\
4 & NFW & $-0.50\pm 0.18$ & $\mathbf{3.0}$ & $\mathbf{7.7}$ & $2.7\pm 0.4$ &
 $\mathbf{2.2}$ & $\mathbf{8}$ & $13.8$ & $10$ \\
5 & NFW & $\mathbf{-1.0}$ & $2.5\pm 0.1$ & $\mathbf{7.7}$ & $2.9\pm 0.4$ &
 $\mathbf{2.2}$ & $\mathbf{8}$ & $10.1$ & $10$ \\
\multicolumn{10}{c}{Bright LRG sample} \\
6 & SIS & $\mathbf{-2.0}$ & $\mathbf{2.0}$ & $\mathbf{10^4}$ & $5.6\pm 0.6$
 & $\mathbf{3.7}$ & $\mathbf{13}$ & $28.6$ & $10$ \\
7 & Power-law & $\mathbf{-2.0}$ & $1.85\pm 0.05$ & 
 $\mathbf{10^4}$ & $7.8\pm 1.1$ & $\mathbf{3.7}$ & $\mathbf{13}$ &
 $16.7$ & $10$ \\ 
8 & NFW & $\mathbf{-1.0}$ & $\mathbf{3.0}$ & $4.7\pm 0.6$ & $6.7\pm 0.8$ &
 $\mathbf{3.7}$ & $\mathbf{13}$ & $10.5$ & $10$ \\
9 & NFW & $-0.72\pm 0.12$ & $\mathbf{3.0}$ & $\mathbf{6.7}$ & $6.4\pm 0.8$
 & $\mathbf{3.7}$ & $\mathbf{13}$ & $11.9$ & $10$ \\
10 & NFW & $\mathbf{-1.0}$ & $2.7\pm 0.1$ & $\mathbf{6.7}$ & $6.8\pm 0.8$ &
 $\mathbf{3.7}$ & $\mathbf{13}$ & $9.4$ & $10$ \\
\hline\hline
\end{tabular}
\end{table*}

\subsubsection{Faint LRG sample}

\begin{figure}
\includegraphics[angle=0,width=3.2in]{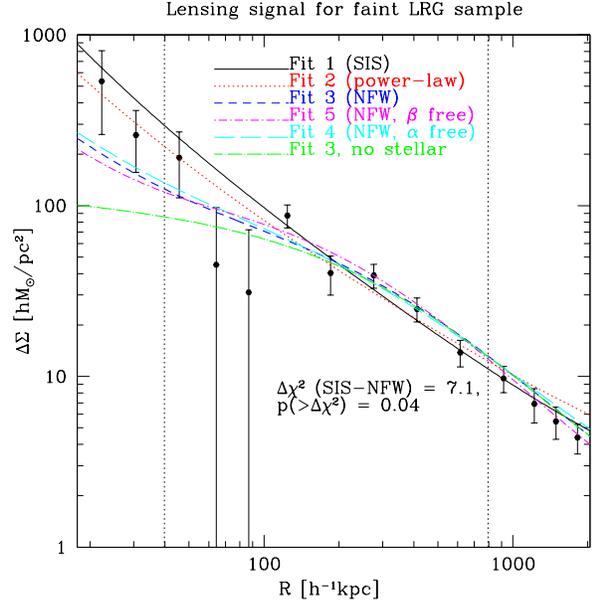}
\caption{\label{F:sigfitfaint}Lensing signal for the faint LRG lens
  sample averaged over source
  sample, and best-fit signals.  The vertical dotted line at 40 \hkpc{} shows
  the minimum scale used for the fits, and at $800$ \hkpc{} shows
  $r_{vir}$.  The line types for each fit  
  are indicated on the plot itself.  We note that the signal shown
  here has been corrected for errors in the boost factor due to
  the sky subtraction error, but non-weak shear
  effects and magnification bias effecs on the boost factor are still
  incorporated in both the real and the model signal shown here.
  Errorbars are slightly correlated (20--30 per cent level) on large scales.} 
\end{figure}

Accordingly, we begin by discussing the results for the faint LRG
sample, with the simplest fit performed, that for a simple SIS with
fixed stellar component, with only one free parameter.
This fit (fit number 1 in Table~\ref{T:fitresults}), along with
several of the others to be discussed, appears in Fig.~\ref{F:sigfitfaint} with
the measured signal.   We have forced the profile in Eq.~\ref{E:genprofile} to
approximate a SIS by making $r_s$ very small, so that the
logarithmic slope is equal to $-\beta=-2$ for the entire radial range of
interest.  
It appears from the top panel of Fig.~\ref{F:sigfitfaint} 
that the model overestimates the signal on $40$--$80$ \hkpc\ scales.

Next, we consider the results for a general single power-law, fit 2.
The addition of a single additional parameter, the exponent of the
power-law ($-\beta$), decreases the $\chi^2$ value by $7.1$ to
$11.5$ relative to fit 1, despite the fairly small
change in exponent from $-2$ to $-1.85\pm 0.06$, suggesting that the
addition of this fit parameter is justified.  To properly evaluate the
goodness of this fit relative to that for the SIS 
using the $\Delta\chi^2$ value, we must take into account that it was
calculated using a bootstrap covariance matrix, and as shown in
Appendix~D of \cite{2004MNRAS.353..529H}, the expected
$\chi^2$ values are not drawn from a $\chi^2$ distribution in this
case.  The probability value
associated with exceeding this value of $\Delta\chi^2$ by chance is
$p(>\Delta\chi^2)=0.04$.
On Fig.~\ref{F:sigfitfaint}, it is apparent that this model is preferred
over the SIS because it does not overestimate the signal on $40$--$80$
\hkpc{} scales as much as the SIS (though the model signal is slightly high
on $1$--$2$ \hmpc{} scales).  

We then consider the simplest NFW-type fit, number
3, which has the same number of fit parameters 
as the single power-law, and the same $\chi^2$. Consequently, like the
single power-law profile, the NFW model is preferred relative to the
SIS at the 96 per cent CL.\footnote{We have used the $\Delta\chi^2$
  values from a distribution 
  assuming $200$ bootstrap regions, $12$ radial bins, and $2$ fit
  parameters.  Since the SIS is not a special case of NFW, the actual
  distribution of $\Delta\chi^2$ is more complicated in a way that is
  difficult to account for analytically, but it implies
  that the value of 96 per cent is overly conservative.}
Visually, Fig.~\ref{F:sigfitfaint} suggests that particularly on $40$--$80$
\hkpc{} and $1$--$2$ \hmpc{} scales, the NFW model is preferred.  The
best-fit value of mass, $(2.9\pm 0.4) \times 
10^{13}h^{-1}M_{\odot}$.
This should be compared to the median mass of the sample
\cite{2005MNRAS.362.1451M}.  It is lower than the value
expected naively from the halo mass function using high normalization, in
Table~\ref{T:dndmresults}; it is close to the predicted lower mass
limit of the sample, $2.5\times 
10^{13}h^{-1}M_{\odot}$, or to the expected mass value using the lower
normalization model.  (Actually placing constraints on $\Omega_m$ and
$\sigma_8$ using abundances is a more involved topic that we will
address in a future publication.) 

We can also compare concentration parameter fits to theoretical expectations. 
Scatter in the mass-concentration relationship can
lower the value in 2-d lensing fits when compared to average in 3-d.  When
using the signal for central galaxies in the brightest luminosity bin 
in the simulations from \cite{2005MNRAS.362.1451M}, which incorporate
both scatter in the mass-luminosity relationship and in the
concentration-mass relationship, we find that the best-fit
concentration is about 20 per cent lower than that expected for the
corresponding mass.  If we assume the same level of underestimate
here, then this means that the fitted values should be increased
by 20 per cent. We use the same value as an estimate of systematic 
error associated with this procedure, but 
this source of systematic error could 
be reduced once large enough samples of simulated clusters are used 
to test the procedure. 
Thus the best-fit value of concentration is
$5.0\pm 0.6 \mbox{(stat)} \pm 1 \mbox{(sys)}$, which should be compared to 
7.7 for the high normalization cosmological model
and 6.5 for the low normalization model. 
We see that there is some tension with the high normalization cosmology, 
but the method needs to be tested further before we can 
draw conclusions about cosmology.

For fit 4, we fix $\beta$ to its canonical value of $3.0$, and $c$ to
$7.7$, but allow
$\alpha$ and mass to vary.  We find that $\alpha$ tends
towards the maximum permitted value in the fits, $-0.5$, or a very
shallow inner 
profile.  This fit has a $\chi^2$ of $13.8$, compared to fit 
3, which had $\alpha$ fixed to the canonical value of $-1.0$ but
allowed $c$ to vary, with a $\chi^2$ of $11.5$.  This small  difference in
$\chi^2$ suggests that
there is little difference between these models, and that we are thus
not highly sensitive to some combination of the values of $c$ and
$\alpha$. 

For fit 5, we instead fix $\alpha$ to its canonical value of $-1.0$,
and $c$ to $7.7$, but allow $\beta$ and mass to vary.
$\beta$ tended 
to suggest a slightly shallower outer asymptotic slope, $-2.5\pm 0.1$
rather than $-3.0$, but still significantly different from the SIS
value of $-2.0$.  Again in comparison with fit 3, the $\chi^2$ for
this fit is roughly $1$ lower, which means this fit is slightly
better, but not in a way that is statistically significant.  

Next, we tried several other fits, all of which had large degeneracies
and thus none of which are shown in Table~\ref{T:fitresults}.  These
fits include:
\begin{enumerate}
\item A fit for $c$, $M$, and $M_{stellar}$: The best-fit value
  of $M_{stellar}$ was quite close to the assumed value, but with
  little constraining power.  This is not too surprising, since our
  fits use scales $R_{min} \sim 5 r_{deV}$, well beyond the bulk of
  the stellar component.  Nonetheless, it is useful to know that
  significantly different values of $M_{stellar}$ than the assumed
  value are not required by this fit.
\item Simultaneous fits for NFW mass and either inner or outer slope
  along with concentration: we lack the power to constrain both the
  concentration and one of the slopes.  As suggested by a comparison
  of fits 2--4 in Table~\ref{T:fitresults} and
  Fig.~\ref{F:sigfitfaint}, $c$ and $\alpha$ are highly degenerate,
  with even fairly different combination of higher $c$ and shallower
  $\alpha$ allowed with little change to the lensing signal. A
  comparison of fits 3 and 5 shows that the same is true for $\beta$
  and $c$.
\end{enumerate}

Hence, while we are able to 
rule out the SIS model relative to NFW for this lens sample at the 96
per cent CL,
we cannot place strong constraints on all profile parameters with
this dataset. 

\subsubsection{Bright LRG sample}

\begin{figure}
\includegraphics[angle=0,width=3.2in]{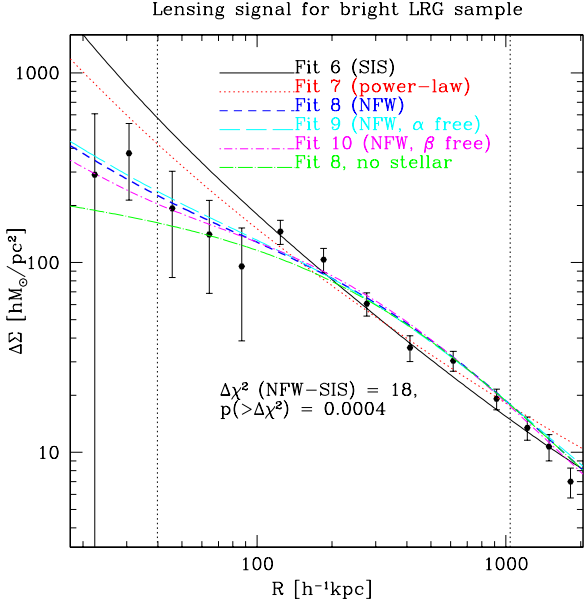}
\caption{\label{F:sigfitbright}Lensing signal for the bright LRG lens
  sample averaged over source
  sample, and best-fit signals.  The dotted line at 40 \hkpc{} shows
  the minimum scale used for the fits, and at $1$ \hmpc\ shows
  $r_{vir}$.  The line types for each fit are indicated on the plot
  itself.  Errorbars are slightly (20--30 per cent level) correlated
  on large scales.} 
\end{figure}
Next, we discuss the fits for the bright LRG sample, fits 6--10 in
Table~\ref{T:fitresults}, with some fits shown with the real signal in
Fig~\ref{F:sigfitbright}.  The first fit we consider is the simple
one-parameter fit (for $M$) with an assumed SIS profile and fixed
stellar components, fit 6.  The
best-fit $\chi^2$ value of $28.6$ gives a 
$p$-value of 0.004, suggesting that this model is a poor fit to the data
From Fig.~\ref{F:sigfitbright}, it is apparent that  the SIS model
once again overpredicts the signal for $40$--$100$ \hkpc{} scales.   

Fit 7 is for a single power-law profile, with the exponent
allowed to vary from the SIS value.  As for the faint LRGs, the
best-fit exponent is $-1.85\pm 0.05$, with a
large decrease in 
best-fit $\chi^2$ of $11.9$ relative to fit 6, and a $p(>\Delta\chi^2)=0.006$,
so the general power-law profile is a markedly improved fit
over the SIS.  Fig.~\ref{F:sigfitbright} indicates that (as for the
faint LRG sample) this is the
case primarily because this model predicts a lower signal than for the
SIS model on $40$--$100$ \hkpc{} scales, at the expense of 
overpredicting the signal on $1$--$2$ \hmpc{} scales.  

Fit 8 is for $c$ and $M$ with fixed NFW $\alpha$, $\beta$, and stellar
components. The best-fit $\chi^2$ is smaller
than that of the SIS model (fit 6) by $18$,
and therefore the NFW model is favored. The SIS
model in fit 6 is ruled out relative to the NFW model at the 99.96 per cent
CL, and the power-law model (fit 7) relative to the NFW at the
94 per cent CL.  As shown in Fig.~\ref{F:sigfitbright}, the
NFW model does a better job predicting the signal on
$40$--$100$ \hkpc{} and $1$--$2$ \hmpc{} scales than either of these
other two models.  This fit suggests a mass of $6.7\pm 0.8 \times
10^{13}h^{-1}M_{\odot}$, so once again smaller than the predicted mean
and median mass of this sample from the high normalization 
model halo mass function in
Table~\ref{T:dndmresults} and closer to the lower limit, or to the
predicted mean and median halo mass using the lower $\sigma_8$ cosmology.  

For the concentration parameter in the NFW profile, we find 
$c=5.6 \pm 0.6 \mbox{(stat)} \pm 1 \mbox{(sys)}$ once we apply the
20 per cent upward correction.  
This should be compared to a predicted $c=6.7$ in the high normalization model 
and $c=5.5$ in low normalization model. While the high normalization 
model predicts somewhat too high values, although within the 
errors, the low normalization model 
agrees well with the observations. 

In fit 9, we constrain the concentration to the predicted
value for this mass of
$6.7$, but allow $\alpha$ to vary.  As shown, the best-fit value is
somewhat shallower than the assumed value of $-1.0$, at roughly the
$3\sigma$ level.  However, the $\chi^2$ value for this fit is worse
than for fit 8 by roughly $1.4$, so the lower value of
concentration is 
 needed to match the signal (but the difference between the two
values of concentration is not statistically significant).

In fit 10, we constrain the concentration to the predicted value of
$6.7$, but allow $\beta$ to vary from canonical NFW value.  As shown,
the best-fit value of $2.7\pm 0.1$ is roughly $3\sigma$ from the
canonical value, and the $\chi^2$ value of $9.4$ is actually slightly
better than fit 18 which has $\beta$ fixed to its canonical value and
$c$ free.  This difference in $\chi^2$ (of $1.1$) is not actually statistically
significant, however. As for the faint LRG sample, we lack the power
to simultaneously contrain concentration along with either the inner
or outer slopes.

For this sample, we also tried fitting simultaneously for $c$, NFW
mass, and $M_{stellar}$.  As for the faint LRG sample, the best-fit
mass was very close to the assumed value, but with very large errors,
so we lack the power to constrain this parameter.  Also as before, we
note that the distinctions between the 
three NFW models shown on Fig.~\ref{F:sigfitbright} is very small
compared to the difference between the SIS, power-law, and NFW models.

\subsubsection{Mass to light relation}\label{SSS:mlrelation}

\begin{figure}
\includegraphics[angle=0,width=3.2in]{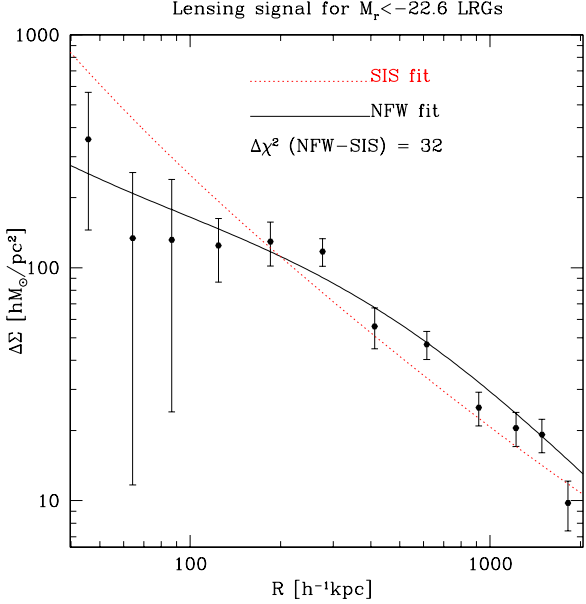}
\caption{\label{F:brightest}Lensing signal for the $M_r<-22.6$ lens
  sample averaged over source sample.} 
\end{figure}

In this section we also include the results of an attempt to determine
the relationship between LRG $r$-band luminosity and mass with finer
binning in luminosity.  The bright bin has been split into two bins at
$M_r=-22.6$.  In this way, we are able to split this sample, which has
an average mass corresponding to massive groups, into one sample with
groups and the other with clusters.  Results for the masses and
concentrations as a function of luminosity appear in
Table~\ref{T:masslum}, and signal for the brightest bin is shown in
Fig.~\ref{F:brightest}.  As shown by this figure, even considering the
small size of this lens sample, we still have significant ability to
constrain the profile and differentiate between NFW and SIS due to the
high halo mass.  When we split the sample further, however, our
ability to determine the profile to high precision degenerates, so we
show no higher 
luminosity bins beyond this.

Concentrations have, 
as in the previous  
section, been adjusted upwards by 20 per cent.  Masses have not been
adjusted to account for scatter in the mass-luminosity relationship,
because the degree of adjustment that is necessary is not yet clear.
In general, however, this adjustment would have to lead to an
increase of the best-fit masses, and hence it is clear that the mass
is increasing quite rapidly with luminosity (roughly $M \propto L^2$).
Thus, the luminosity of the central galaxy can be used as a rough 
measure of the cluster mass, at least for masses below 
$2\times 10^{14}M_{\odot}$, although the (possibly significant)
scatter in this relation  
cannot be established by our statistical approach. This 
approach to finding clusters is much simpler than other methods, such
as the number of red galaxies (i.e. richness) or integrated light within a
certain window in real and redshift space.

\begin{table}
\caption{\label{T:masslum}Masses and concentrations as a function of
  luminosity with finer binning in luminosity.} 
\begin{tabular}{cccc}
\hline\hline
$M_r$ limits & $\langle L/L_{\odot} \rangle$ & $M_{nfw,fit}$ &
$c_{nfw}$ \\
 & [$h^{-2}10^{10}$] & [$10^{13}h^{-1}M_{\odot}$] & \\
\hline\hline
$M_r > -22.3$ & $5.2$ & $2.9\pm 0.4$ & $5.0 \pm 0.6 \mbox{(stat)} \pm
1 \mbox{(sys)}$ \\
$[-22.6, -22.3]$ & $7.6$ & $4.4\pm 0.7$ & $6.8 \pm 1.1 \mbox{(stat)}
\pm 1 \mbox{(sys)}$ \\
$M_r \le -22.6$ & $11.0$ & $12.9 \pm 1.9$ & $5.0 \pm 0.7 \mbox{(stat)}
\pm 1 \mbox{(sys)}$ \\
\hline\hline
\end{tabular}
\end{table}

\begin{figure}
\includegraphics[angle=0,width=3.4in]{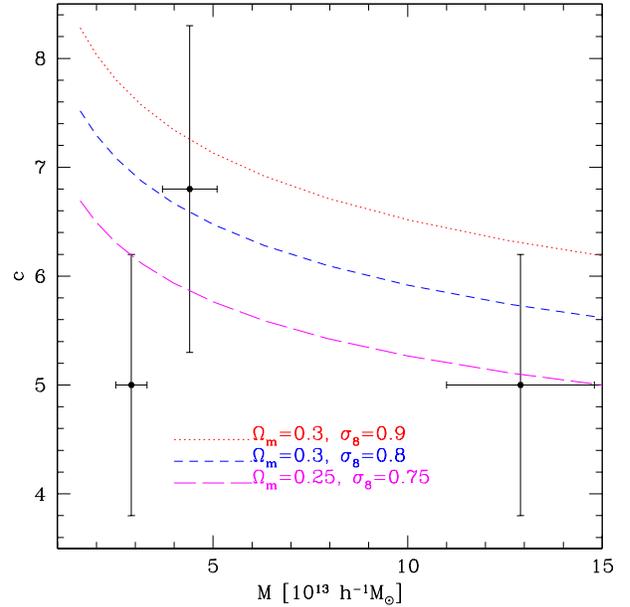}
\caption{\label{F:cmrelation}Measured NFW concentrations with
  statistical and systematic errors added in quadratures, shown as a
  function of best-fit NFW mass.  The lines shown on the plot are
  predictions for several different cosmologies.} 
\end{figure}

Fig.~\ref{F:cmrelation} shows the measured $c$ as a function of halo
mass, along with the predictions for several cosmological models; as
shown, somewhat lower normalizations are preferred.  We
postpone the comparison between
the measured and predicted
abundances to a future publication, 
since more work will be necessary to understand the
effects of scatter in the $M(L)$ relationship on the measured NFW masses.

\section{Interpretation and Conclusions}\label{S:conclusion}

While our fits only use scales above 40 \hkpc\ due to concerns about
lensing systematics,
Figs.~\ref{F:sigfitfaint} and~\ref{F:sigfitbright} include scales down
to 20 \hkpc\, the minimum scale probed.  As shown in
Fig.~\ref{F:sigfitfaint}, the signal below 40 \hkpc\ is fit somewhat
better by the single power-law models than by the NFW profile for the
faint LRG sample, but
this difference is not statistically significant, and can also be
explained by an underestimation of stellar mass rather than by any
problem with the NFW profile.
Fig.~\ref{F:sigfitbright} shows that the NFW model does a satisfactory
job describing the data even below 40 \hkpc\ for the brighter lenses.
For the faint lenses, several changes in parameters could raise the
signal on scales below 40 \hkpc, such as an increase in stellar mass or
a steeper inner slope for the DM profile.  In both figures, we show
the signal for the NFW fit without stellar component, and it is clear
that the inclusion of the stellar component improves the agreement
between model and observation below 50 \hkpc\ scales.

One might conclude, from Figs.~\ref{F:sigfitfaint}
and~\ref{F:sigfitbright}, that it would be difficult to distinguish
between inner DM profiles with any technique.  However, there are
several reasons why this is not the case.  First, in using the lensing
signal to distinguish between these density profiles, we are
necessarily susceptible to some parameter degeneracies that are not 
a problem for other methods, such as those against which we compare
below.  Second, other methods have a different dynamic range; in
particular both strong lensing and kinematic studies can probe to far
smaller scales ($<10$ \hkpc).  When combined with methods such as weak
lensing that can constrain the profile on larger scales, in
conjunction with constraints on the stellar profile, it seems
conceivable that tight constraints could be placed on the inner dark
matter profile. 

We note that the fact that the models are either comparable with or
larger than the lensing signal on $1$--$2$ \hmpc\ scales implies that
we have successfully extracted a sample of LRGs that resides almost
exclusively in host halos rather than as satellites sitting at outskirts 
of clusters for which the signal would be reduced.

There are a number of conclusions that we can draw from these fit
results.  First, the division into stellar and halo components
allows us to fit the data very well.  Adiabatic contraction does not
seem to be necessary for our description on the scales of interest
($R>40$ \hkpc)
because of the relatively low stellar to halo mass ratio, but the
addition of the stellar 
component improves the fit on small scales over the simple NFW profile 
and there is some evidence that it may even be underestimated for the 
faint sample.  We note that it is possible that AC occurred at a
significantly earlier stage in the galaxy's formation, when
$M_{stellar}/M_{halo}$ was much larger.  In that case, more
significant AC on larger scales would 
have caused a steepening of the profile, which would then
approximately be preserved through its later mergers
\citep{2005MNRAS.362..184B,2006ApJ...641..647K}.  This 
is another possible explanation for the excess over the NFW plus
stellar mass model on $20 < R < 40$ \hkpc\ scales; however, due to the
uncertain systematics in this regime, we do not attempt to place
constraints directly.  Furthermore, we are unable to constrain the
form of the stellar component; on the scales under consideration, a
point mass works nearly as well as a Hernquist profile.

We compare these results with those from other methods that are
used to determine the density profiles.  Recent X-ray studies have
found that the profile is consistent with the NFW model \citep{2005astro.ph..7092V,2006ApJ...636..698F,2006astro.ph..1301H},
though with indications for poor groups (consistent in mass with our
``faint'' sample) of some excess over NFW on small scales
\citep{2005astro.ph..7092V}, similar to our finding that the data is
above the NFW model with stellar component in Fig.~\ref{F:sigfitfaint}
for the fainter LRG sample. Thus, our results seem to be consistent
with those from X-ray studies.  

Some kinematic and strong lensing studies (e.g., 
\citealt{2001ApJ...549L..33R},
\citealt{2003ApJ...583..606K}, 
\citealt{2004ApJ...600L...7H}, \citealt{2006astro.ph..1628K}) have suggested  
a logarithmic slope $\sim -2$  on scales below $10$ kpc, where optical 
observations
provide velocity dispersion information and strong lensing occurs. 
It is entirely possible that in that regime
adiabatic contraction actually is necessary to understand the profile,
and leads to the steepening of the profile compared to NFW.  
It is also possible that due to
selection effects in the strong-lensing samples, they are not entirely
comparable to our LRG-selected sample of groups and clusters.
We note that we have central velocity dispersions for the LRG sample used in 
this analysis, which has been used to learn about AC and dark matter 
inside the optical radius \citep{2004NewA....9..329P}. 
In the future it would be interesting to 
combine the weak lensing information with the kinematic information to 
learn more about the mass profile in the inner parts of the clusters. 

Studies of galaxy kinematics on larger scales, out to several Mpc,
have found a strong preference for the NFW profile over a SIS profile
on the same scales as this study.  These works (e.g.,
\citealt{2000AJ....119.2038V}; \citealt{2003ApJ...585..205B};
\citealt{2003AJ....126.2152R}; \citealt{2006astro.ph..2032R})
typically focus on several tens of clusters with a large number of
spectroscopic redshifts per cluster, and determine the profile using
the cluster infall patterns.  Generally the clusters for which this
technique is used are in a higher mass range than those considered
here, $10^{14}$ - $10^{15} h^{-1}M_{\odot}$.  It is useful to know
that our results, which focus on the average profile of groups and
clusters in a lower mass range, have been found to be true for
individual higher mass clusters as well.

We can see from the fits to the two samples
separately that a characteristic halo mass for LRGs overall is
$\sim 5\times 10^{13} h^{-1}M_{\odot}$.  This corresponds to the transition 
regime between groups and low mass clusters.
This is not too
surprising, since the comoving number density of these objects
($1.2\times 10^{-4} (h/\mbox{Mpc})^3$) is an order of magnitude larger than
for large clusters, and hence the typical LRG halo mass
must be significantly smaller than that of massive clusters.  
If we adopt the definition that clusters have virial masses above 
$10^{14} M_{\odot}$ then, given that the best fit mass of our bright sample 
corresponds to this value when we use $h=0.7$, half of the halos in the
bright sample can be defined as low mass clusters. 
This is also
consistent with the findings in Loh \& Strauss (2006, {\it in prep.})
that the typical LRG resides in groups, with some in the field, and
some fraction in large clusters.
We find that on average, brighter central LRGs reside in more massive
halos, and the relation scales as $M \propto L^2$ over the range we observe. 

When fitting to 2 \hmpc, a profile that has a change in logarithmic
slope with scale 
seems to provide the best fit to the data; the SIS profile is ruled
out more robustly than a general power-law profile.  Specifically, for
the faint LRG sample, the SIS is ruled out relative to the NFW profile
at the 96 per cent CL;
for the bright LRG sample, the SIS is
ruled out relative to NFW at the 99.96 per cent CL.


 Use of
the the canonical slopes $\alpha=-1$ and $\beta=3$ gives best-fit
concentrations of $5.0\pm 0.6 \pm 1$ and $5.6\pm 0.5 \pm 1$ for faint
and bright 
lens samples, respectively, where the first error is statistical and 
second systematic, since the scatter in concentration-mass
relation leads to a bias in the fits and we do not have sufficiently large 
simulated samples of groups and clusters to model this bias
accurately. By splitting the brighter bin further, we are able to
obtain two different concentration measurements for group and
cluster-sized halos individually, as discussed in \S\ref{SSS:mlrelation}.
The results are lower than the expected values from high normalization 
$\Lambda$CDM simulation, as shown in Fig.~\ref{F:cmrelation}.
If we allow $\Omega_m=0.25$ and $\sigma_8=0.75$ in accordance with the
recent WMAP 3-year results \citep{2006astro.ph..3449S}, the predicted
values of $c$ are low enough that the discrepancies are reduced to an
acceptable level (below $1\sigma$ when both statistical and systematic
errors are included). Note that in principle, the statistical 
errors are sufficiently small that the method can provide 
constraining power to differentiate among currently viable cosmological 
models. We refrain from making strong cosmological 
implications from this comparison because the method has not been 
calibrated yet in simulations over this halo mass range. When 
applied to lower masses below $10^{13}M_{\odot}$ it was found that 
the method underestimates the true concentration parameter by about 
20 per cent \citep{2005MNRAS.362.1451M}, 
a correction we have applied in the present analysis as well. 

In general, therefore, our results are in accordance with the
predictions of $\Lambda$CDM N-body simulation predictions of halo
profiles and statistically the measurements have very high signal to noise. 
More detailed analysis in the future will be 
able to differentiate among the cosmological models using both
the measured cluster profiles and the abundance information.

\section*{Acknowledgments}

U.S. is supported by the Packard Foundation, NASA NAG5-1993 and NSF 
CAREER-0132953. R.C. acknowledges funding from a National Science
Foundation Graduate Research Fellowship.  C.H. is supported in part by
NSF PHY-0503584 and by a  
grant-in-aid from the W. M. Keck Foundation.  We thank Jim Gunn, Robert 
Lupton, Nikhil Padmanabhan, Yeong-Shang Loh, David Wake, Erin
Sheldon, and Chung-Pei Ma for useful conversations.

Funding for the SDSS and SDSS-II has been provided by the Alfred
P. Sloan Foundation, the Participating Institutions, the National
Science Foundation, the U.S. Department of Energy, the National
Aeronautics and Space Administration, the Japanese Monbukagakusho, the
Max Planck Society, and the Higher Education Funding Council for
England. The SDSS Web Site is http://www.sdss.org/. 

The SDSS is managed by the Astrophysical Research Consortium for the
Participating Institutions. The Participating Institutions are the
American Museum of Natural History, Astrophysical Institute Potsdam,
University of Basel, Cambridge University, Case Western Reserve
University, University of Chicago, Drexel University, Fermilab, the
Institute for Advanced Study, the Japan Participation Group, Johns
Hopkins University, the Joint Institute for Nuclear Astrophysics, the
Kavli Institute for Particle Astrophysics and Cosmology, the Korean
Scientist Group, the Chinese Academy of Sciences (LAMOST), Los Alamos
National Laboratory, the Max-Planck-Institute for Astronomy (MPA), the
Max-Planck-Institute for Astrophysics (MPIA), New Mexico State
University, Ohio State University, University of Pittsburgh,
University of Portsmouth, Princeton University, the United States
Naval Observatory, and the University of Washington.

\bibliography{../BibTeX/apjmnemonic,../BibTeX/cosmo,../BibTeX/cosmo_preprints}
\bibliographystyle{mn2e}

\appendix

\section{Effects of non-weak shear on shear estimator}\label{A:nonweak}

Here we compute the relation between the mean ellipticity and the 
distortion matrix.  This relation has previously been examined by 
\citet{1995A&A...294..411S} for the case with $\omega=0$; they derived an 
equation that is essentially equivalent to our Eq.~(\ref{eq:C:em}), 
although they did not provide the Taylor expansion in $g$.  The analysis 
in this paper only makes use of the terms through second order in the weak 
lensing approximation (i.e. second order in $\kappa$ or in 
$\Sigma_c^{-1}$) since this is adequate for the LRGs at tens of kpc radii 
where $\kappa\sim 0.1$.  We do however present the full formalism here, 
including all orders in $\Sigma_c^{-1}$, in the hope that it will be 
useful in the future.

The analysis here makes extensive use of the fact that our 
ellipticities transform covariantly under shear (this is in fact true for 
all the adaptive ellipticities, e.g. as considered by 
\citealt{2002AJ....123..583B}); note that many of the other ellipticities 
used in the literature, such as ``KSB'' ellipticities 
\citep{1995ApJ...449..460K}, do not share this property and the results of 
this section do not apply to them.  

The local relation between the source and image planes is usually
described by the Jacobian: 
\beq
{\MJ} = \frac{\partial {\bmath x}_S}{\partial {\bmath x}_I}
=  \left(\begin{array}{cc} 1-\gamma_+-\kappa & -\gamma_\times+\omega
    \\ -\gamma_\times-\omega & 1+\gamma_+-\kappa\end{array} \right), 
\eeq
where the rows indicate the $1$ and $2$ components of ${\bmath x}_S$
and the columns indicate those of ${\bmath x}_I$.  Our basic problem
here is to  
relate the mean ellipticity of the galaxies, $\langle{\bmath
  e}\rangle$, to the entries in $\MJ$.  In particular, we will find
that  
$\langle{\bmath e}\rangle$ is not simply proportional to $\bgamma$.

It is most convenient here to work with the singular-value
decomposition of $\MJ$ rather than with $\MJ$ itself.   
The singular-value decomposition is $\MJ =
\MR(\varphi)\MD\MR(\theta)$, where $\MD$ is diagonal and $\MR(\alpha)$
represents a counterclockwise  
rotation by angle $\alpha$.  If we define ${\cal A}={\rm Tr}\,\MD/2$,
then this can be written as 
\beq
\MJ = {\cal A}\MR(\varphi)\
\left(\begin{array}{cc} 1-g & 0 \\ 0 & 1+g \end{array}\right)
\MR(\theta)
\equiv
 {\cal A}\MR(\varphi)\MT(\delta)\MR(\theta).
\label{eq:C:j}
\eeq
The four variables $\{\gamma_+,\gamma_\times,\kappa,\omega\}$ in the
Jacobian have thus  
been replaced with the four variables $\{g,\theta,{\cal A},\varphi\}$.
The explicit formula for the conversion can be derived from the  
rectangular-to-polar conversion, Eq.~(A5) of
\citet{2003MNRAS.343..459H}, which gives: 
\beqa
2(1-\kappa) &=& 2{\cal A}\cos(\varphi+\theta), \nonumber \\
-2\omega &=& 2{\cal A}\sin(\varphi+\theta), \nonumber \\
-2\gamma_+ &=& -2g\cos(\varphi-\theta), {\rm ~and} \nonumber \\
-2\gamma_\times &=& -2g\sin(\varphi-\theta).
\eeqa
The solution to this system of equations is
\beqa
{\cal A} &=& \sqrt{(1-\kappa)^2+\omega^2}, \nonumber \\
g &=& \sqrt{\frac{\gamma_+^2+\gamma_\times^2}{(1-\kappa)^2+\omega^2}},
\nonumber \\ 
\varphi &=& \frac{1}{2}\left( -\arctan\frac{\omega}{1-\kappa} +
  \arctan\frac{\gamma_\times}{\gamma_+}\right), 
{\rm ~and} \nonumber \\
\theta &=& \frac{1}{2}\left( -\arctan\frac{\omega}{1-\kappa} -
  \arctan\frac{\gamma_\times}{\gamma_+}\right). 
\eeqa
The arguments of $\arctan$ in the last two lines have been written
such that the numerator and denominator give the correct quadrant for
the  
arctangent.  Note that for the case of a single-screen lens with
$\omega=0$, $g$ reduces to the familiar ``reduced shear''
$g=\gamma/(1-\kappa)$ 
\citep{1995A&A...294..411S}.

Of these four variables, ${\cal A}$ re-scales the source plane and so
has no effect on the ellipticities.  The rotation $\MR(\varphi)$
rotates the  
source plane and for randomly oriented source galaxies can have no
effect on the average ellipticity.  The transformation $\MT(g)$ in  
Eq.~(\ref{eq:C:j}) can produce net ellipticity in the $+$ direction;
and $\MR(\theta)$ rotates the image plane by angle $\theta$ and hence  
$\langle\bmath e\rangle$ by angle $2\theta$.  Thus the observable
$\langle\bmath e\rangle$ depends only on $g$ and $\theta$, and
moreover the  
dependence on $\theta$ corresponds trivially to a rotation.

Our problem is thus reduced to determining $\langle\bmath e\rangle$
for the special case where $\MJ=\MT(g)$.  In this case symmetry
considerations  
imply $\langle e_\times\rangle=0$, so we need only determine $\langle
e_+\rangle$.    If the intrinsic ellipticity of a galaxy is
$\bmath\eint$, then  
the equation for the final ellipticity is given by Eq.~(2.5) of
\citet{1991ApJ...380....1M}: 
\beq
e_+ = \frac{\eint_+ + \delta}{1+\delta \eint_+},
\eeq
where the distortion $\delta$ is related to $g$ by Eqs.~(2-7) and
(2-8) of \citet{2002AJ....123..583B}: 
\beq
\delta = \tanh( 2\arctanh g) = \frac{2g}{1+g^2}.
\label{eq:C:gdelta}
\eeq
It is simplest to evaluate $\langle e_+\rangle$ for a population of
sources whose intrinsic ellipticities have fixed magnitude $\eint$ and
random  
orientation $\phi$; other source populations can be handled by further
averaging over the appropriate distribution for $\eint$.  In this
case, we may  
write $\eint_+=\eint\cos 2\phi$, so that
\beqa
\langle e_+\rangle &=& \frac1{2\pi} \int_0^{2\pi} \frac{\eint\cos
  2\phi + \delta}{1+\delta \eint\cos 2\phi} \,\rmd\phi 
\nonumber \\
&=&\frac1{2\pi}\int_0^{2\pi} \left(\delta^{-1} +
  \frac{\delta-\delta^{-1}}{1+\delta \eint\cos 2\phi}\right)
\,\rmd\phi 
\nonumber \\
&=& \delta^{-1} + \frac{\delta-\delta^{-1}}{\sqrt{1-\delta^2\eints{2}}}.
\label{eq:C:em}
\eeqa
(The first term in the second line is integrated trivially, while the
second term can be reduced to a rational integral by the substitution  
$z=\tan\phi$.)  We are interested in the weak-shear regime where we
can use the Taylor expansion in $\delta$, which is easily obtained
from the  
binomial expansion of $(1-\delta^2\eints{2})^{-1/2}$:
\beqa
\langle e_+\rangle &=& \left( 1 - \frac12\eints2 \right)\delta
+ \left( \frac12\eints2 - \frac38\eints4 \right)\delta^3
\nonumber \\ &&
+ \left( \frac38\eints4 - \frac5{16}\eints6\right)\delta^5
\nonumber \\ &&
+ \left( \frac5{16}\eints6 - \frac{35}{128}\eints8\right)\delta^7
+ ...\;.
\label{eq:C:taylor1}
\eeqa
(There is no singularity at $\delta=0$ because the coefficients of
$\delta^{-1}$ cancel.) 
In the weak lensing literature, $g$ is usually used instead of
$\delta$; substitution of Eq.~(\ref{eq:C:gdelta}) into
Eq.~(\ref{eq:C:taylor1}) yields 
\beq
\langle e_+\rangle = \sum_{n=0}^\infty \zeta_ng^{2n+1},
\label{eq:C:taylor2}
\eeq
where the first few coefficients in the expansion are
\beqa
\zeta_0 &=& 2-\eints2,
\nonumber \\
\zeta_1 &=& -2+5\eints2-3\eints4,
\nonumber \\
\zeta_2 &=& 2-13\eints2+21\eints4-10\eints6, {\rm ~and}
\nonumber \\
\zeta_3 &=& -2+25\eints2-78\eints4+90\eints6-35\eints8.
\eeqa
For an ensemble of source galaxies with different intrinsic
ellipticities, the values $\eints n$ should be replaced by the
appropriate moments  
$\langle\eints n\rangle$ of the source galaxy distribution.

In the case of a single-screen lens, it is convenient to expand
$\langle e_+\rangle$ in powers of the lensing strength
$\Sigma_c^{-1}$, since both  
$\gamma$ and $\kappa$ are proportional to this quantity.  In
particular, we have 
\beq
\gamma_+ = \Delta\Sigma_+\,\Sigma_c^{-1}, \quad
\gamma_\times = \Delta\Sigma_\times\,\Sigma_c^{-1},
{\rm ~and~} \kappa = \Sigma\,\Sigma_c^{-1}.
\eeq
Here $\Sigma$ is the surface density at the location of the image and
$\Delta\bmath\Sigma$ is the surface density contrast defined in
Fourier space by  
\beq
\Delta\Sigma_+(\bmath l) \pm \rmi \Delta\Sigma_\times(\bmath l) =
\rme^{\pm2\rmi\arctan(l_2/l_1)} \Sigma(\bmath l). 
\eeq
In the coordinate system where $\Delta\bmath\Sigma$ is in the $+$
direction, Eq.~(\ref{eq:C:taylor2}) then becomes 
\beq\label{E:sumn}
\langle e_+\rangle = \sum_{n=0}^\infty \zeta_n \left(
  \frac{|\Delta\bmath\Sigma|\,\Sigma_c^{-1}}{1-\Sigma\,\Sigma_c^{-1}}
\right)^{2n+1}. 
\eeq
This is turned into a power series in $\Sigma_c^{-1}$ using the
binomial theorem for negative powers, 
\beq
(1-x)^{-(2n+1)} = \sum_{j=0}^\infty \combo{2n+j}{2n} x^j,
\eeq
which gives:
\beq
\langle e_+\rangle = \sum_{n=0}^\infty \zeta_n
|\Delta\bmath\Sigma|^{2n+1}\Sigma_c^{-(2n+1)} 
  \sum_{j=0}^\infty \combo{2n+j}{2n} \Sigma^j\Sigma_c^{-j}.
\eeq
We now consolidate terms with the same exponent by defining $k=2n+j+1$:
\beqa\label{E:sumk}
\langle e_+\rangle &=& \sum_{k=1}^\infty 
\left[
\sum_{n=0}^{\lfloor (k-1)/2\rfloor}
\combo{2n+j}{2n}
\zeta_n |\Delta\bmath\Sigma|^{2n+1} \Sigma^{k-2n-1}
\right]
\nonumber \\ && \times \Sigma_c^{-k}.
\eeqa

For the purposes of this work, we only use the $k=1$ term in
Eq.~\ref{E:sumk} (or equivalently, the $n=0$ term in
Eq.~\ref{E:sumn}), since higher order terms are 3rd-order or higher in
$\kappa$ and $\gamma$.  This lowest order term gives
\beq
\langle e_+\rangle = (2-e^{(0)2}) \frac{|\Delta\bmath\Sigma|\,\Sigma_c^{-1}}{1-\Sigma\,\Sigma_c^{-1}}
\eeq
we identify the $(2-e^{(0)2})$ factor with $2{\cal R}$, twice the
shear responsivity.  We then do the lowest order Taylor expansion of
this equation, which tells us that
\beq
\langle e_+ \rangle = 2{\cal R} \langle \Delta\Sigma\,\Sigma_c^{-1} + \Delta\Sigma\,\Sigma\,\Sigma_c^{-2}\rangle
\eeq

When we do the usual approximation that the lenses are statistically
identical, which is reasonable for these volume-limited samples, then
the correction due to non-weak shear is
\beq\label{E:nonweak}
\langle e_+ \rangle = 2{\cal R} \langle\Delta\Sigma\rangle
\langle\Sigma_c^{-1}\rangle \left( 1 +
  \frac{\langle\Delta\Sigma\,\Sigma\rangle}{\langle\Delta\Sigma\rangle}\frac{\langle \Sigma_c^{-2} \rangle}{\Sigma_c^{-1}}\right)
\eeq

Note that the factor in parenthesis is not the same as $1+\kappa$,
for two reasons.  First,
$\langle\Delta\Sigma\,\Sigma\rangle/\langle\Delta\Sigma\rangle >
\langle\Sigma\rangle$ since the massive objects get weighted more
heavily.  Second, $\langle\Sigma_c^{-2}\rangle /
\langle\Sigma_c^{-1}\rangle > \langle\Sigma_c^{-1}\rangle$.  Both of
these issues will tend to make reduced shear more important than the
naive argument in the text that it depends on the size of $1+\kappa$.

\end{document}